\documentclass[%
 reprint,
 amsmath,amssymb,
 aps,
 prapplied,
 floatfix,
]{revtex4-2}

\usepackage{graphicx}
\usepackage{dcolumn}
\usepackage{bm}
\usepackage{amsmath}
\usepackage{amssymb}
\usepackage{booktabs}
\usepackage{hyperref}

\begin{document}

\title{Identifying the Shear Piezoelectric Coefficient using Physics-Informed Neural Networks under Exact Boundary and Initial Constraints}

\author{Suhas Suresh Bharadwaj}
\email{f20230029@dubai.bits-pilani.ac.in}
\affiliation{Department of Electrical and Electronics Engineering\\
 BITS Pilani, Dubai Campus, Dubai, United Arab Emirates}

\date{\today}

\begin{abstract}
The shear piezoelectric coefficient $e_{15}$ governs the shear-mode electromechanical coupling in piezoelectric micromachined ultrasonic transducers (PMUTs) and surface acoustic wave (SAW) devices, yet its measurement typically requires destructive resonance testing on dedicated specimens. This paper shows, on simulated data, that $e_{15}$ can instead be recovered non-destructively from sparse, noisy subsurface wavefield observations using a constrained physics-informed neural network (PINN). By building exact spatial distance functions and temporal multipliers directly into the network output, the clamped, grounded boundary conditions and the resting initial state are satisfied automatically, leaving the loss function with only the interior equations of coupled electro-elastodynamics and the data misfit. Along with a non-dimensional scaling that brings all governing equation coefficients within a single decade of $\mathcal{O}(1)$, this setup removes the boundary and initial-condition penalty weights that soft-constraint methods have to tune. In a six-way comparison on a plane-strain PZT-5H problem with a manufactured wavefield and 30~dB noise, a soft-penalty baseline with adaptive weights converges to an erroneous $e_{15}$ in four of the five runs ($51.7\%$ mean error), and fixed penalties reach an error of $0.85\%$ only after a second-order optimization stage. In contrast, the constrained formulations proposed here reach mean $e_{15}$ errors between $0.15\%$ and $0.51\%$ with held-out displacement errors below $1\%$. Within these variants, the spatial multiplier gives rapid first-order convergence, and the quadratic temporal multiplier lowers the early-time electrostatic residual by a factor of $6.8$ compared with a linear one. Tested on finite-element data along a single subsurface line, the full formulation recovers $e_{15}$ with a mean error of $1.19\%$ at 30~dB and keeps the mean error between $0.65\%$ and $2.71\%$ across a noise sweep from 40~dB down to 0~dB. Finally, since the internal electric potential cannot be measured in practice, training on mechanical displacements alone still recovers $e_{15}$ with mean errors of $0.45\%$ on finite-element data and $1.64\%$ on manufactured data.
\end{abstract}

\maketitle

\section{Introduction}
\label{sec:intro}

Characterizing the electromechanical coupling tensor of anisotropic thin films typically requires destructive resonance testing. For transversely isotropic media such as Lead Zirconate Titanate (PZT) and Scandium Aluminum Nitride (ScAlN) \cite{Tasnadi2010}, the diagonal piezoelectric stress components are easy to extract from standard thickness-stretch and radial resonant modes \cite{auld1973acoustic, 26560}. The shear coefficient $e_{15}$ is harder. It controls how efficiently lateral shear waves are excited and sets the shear-horizontal electromechanical coupling, and it is conventionally isolated through pure shear resonances in specifically diced, laterally poled samples \cite{tiersten1969linear}. Inverse procedures that match simulated and measured electrical impedance can identify complete sets of piezoelectric parameters \cite{claes2025machine}, but they also work on dedicated specimens and not on embedded or in-situ structures. This limits quality control during the fabrication of piezoelectric micromachined ultrasonic transducers (PMUTs) \cite{Muralt2008} and surface acoustic wave (SAW) devices \cite{Delsing2019}, where wafer-to-wafer variation in $e_{15}$ changes the resonant frequency, the coupling efficiency, and the bandwidth of the finished transducer. Embedded subsurface sensing, such as fiber Bragg grating arrays in structural composites \cite{glisic2007fibre, kuang2001embedded}, shows that internal fields can be interrogated without destructive access. This paper therefore tests whether $e_{15}$ can be recovered from sparse, one-sided, subsurface observations instead of full-field measurements, which would allow in-situ, non-destructive characterization.

Existing NDE approaches for thin-film piezoelectric characterization include laser interferometry \cite{kholkin1996interferometric}, electrical impedance spectroscopy \cite{barsoukov2018impedance}, and Brillouin light scattering \cite{carlotti1994brillouin}. While these methods can resolve certain elastic and piezoelectric constants with high accuracy, isolating $e_{15}$ non-destructively is generally difficult, since its signature in the surface displacement spectrum is coupled to, and can be partially masked by, the dominant longitudinal constants ($c_{33}$, $e_{33}$). Decoupling $e_{15}$ from an acoustic measurement therefore requires solving a nonlinear inverse parameter identification problem over a strongly coupled, anisotropic, stiff transient wavefield \cite{drain2019laser, royer1999elastic}.

Finite element (FE) based inversion \cite{allik1970finite} combined with adjoint-state gradients provides a well-established route to full-waveform inversion \cite{virieux2009fwi, tarantola2005inverse}, but each iteration requires forward and adjoint wavefield solves over many time steps, and the cost grows with mesh density and with the number of unknown fields. Mesh-free surrogate models have therefore been explored as alternatives, particularly when several fields must be inferred jointly from sparse observations \cite{markidis2021old}. In the present work, the recovery of a single coefficient serves as a controlled test bed for the constrained architecture, which accommodates further unknown coefficients as additional trainable parameters.

Scientific Machine Learning (SciML) offers a mesh-free alternative for problems of this kind. Physics-Informed Neural Networks (PINNs) \cite{raissi2019physics, karniadakis2021physics, cuomo2022scientific} write the governing partial differential equations (PDEs) into the training objective of a neural network and compute the required derivatives by reverse-mode Automatic Differentiation (AD) \cite{baydin2018automatic}. They have been applied to forward and inverse problems in fluid dynamics \cite{cai2021physics}, linear elastodynamics \cite{haghighat2021physics, rao2021physics}, and acoustic NDE \cite{shukla2020physics}. For piezoelectric systems, PINNs have been used for plate modelling \cite{tandis2025physics} and shunted piezoelectric systems \cite{mouratidou2026physics}, but, to the author's knowledge, they do not recover the shear coefficient $e_{15}$ from sparse transient wavefield data.

Despite these successes, standard PINN formulations often struggle with stiff, strongly coupled multiphysics problems such as piezoelectricity, largely because the physical scales of the mechanical and electrical terms differ so much. In standard continuous PINNs, for example those built with the DeepXDE framework \cite{lu2021deepxde}, the training objective is a sum of the interior PDE residuals and penalty terms for the boundary conditions (BC) and initial conditions (IC):

\begin{equation}
\mathcal{L}(\theta) = \mathcal{L}_{\mathrm{PDE}}(\theta)
  + \lambda_{\mathrm{BC}}\,\mathcal{L}_{\mathrm{BC}}(\theta)
  + \lambda_{\mathrm{IC}}\,\mathcal{L}_{\mathrm{IC}}(\theta),
\label{eq:soft_loss}
\end{equation}

where $\theta$ represents the network weights and $\lambda$ are penalty hyperparameters. Wang et al. \cite{wang2021understanding} and Krishnapriyan et al. \cite{krishnapriyan2021characterizing} showed that enforcing boundary conditions through soft penalties can create gradient-flow pathologies in stiff systems, and piezoelectricity sits toward the extreme end of that spectrum. The elastic stiffnesses ($c^{E}_{ij} \sim 10^{10}$ Pa) and the dielectric permittivities ($\varepsilon^{S}_{ij} \sim 10^{-8}$ F/m) differ by many orders of magnitude as SI numbers, and the mechanical and electrical residuals carry different units (N/m$^3$ and C/m$^3$), so in an unscaled mean squared error their raw magnitudes can be many orders apart. Wang et al. analyzed imbalances of this kind for general PINNs through gradient statistics \cite{wang2021understanding} and Neural Tangent Kernel (NTK) eigenvalues \cite{wang2022and}, and a comparable imbalance is expected here.

A common remedy is self-adaptive loss weighting or NTK-based balancing, which adjusts the $\lambda$ values during training \cite{braga2021self, wang2022and, bischof2025multi}. These methods help in many problems, but Sec.~\ref{sec:baseline} shows that a gradient-norm version of them is not reliable for the inverse problem studied here. Soft penalties also have a second weakness in this problem. The boundaries are only approximately satisfied early in training, so the simulated wave picks up phase errors, and one possible consequence is that the optimizer lowers the data loss by pushing $e_{15}$ upward instead of fixing the boundaries. Through piezoelectric stiffening, $c_{44}^D \approx c_{44}^E + e_{15}^2/\varepsilon_{11}$ \cite{auld1973acoustic}, a larger $e_{15}$ raises the effective shear wave speed and could bring the phase back into line at the observation plane, which would resemble the deceptive convergence described by Krishnapriyan et al. \cite{krishnapriyan2021characterizing}. Sec.~\ref{sec:baseline} treats this as a hypothesis that the experiments support but do not prove. The alternative is to enforce the boundaries exactly instead of through penalties. This goes back to Lagaris, Likas, and Fotiadis \cite{lagaris1998artificial}, who showed that trial solutions satisfying the boundary conditions by construction reduce the optimization to the interior of the domain.

Sukumar and Srivastava \cite{sukumar2022exact} later showed that distance functions can impose spatial boundary conditions exactly on general geometries, and hard enforcement of initial and boundary conditions has also been used for transient elastodynamics \cite{rao2021physics}. The present work builds on these ideas and on a one-dimensional electro-elastodynamic formulation with exact geometric constraints reported earlier by the author \cite{bharadwaj2026thermodynamically}. It extends them to a two-dimensional coupled hyperbolic-elliptic piezoelectric system, examines how the quasi-static potential behaves near $t=0$, and, to the author's knowledge, uses a hard-constrained network for the first time to identify a piezoelectric coefficient, $e_{15}$, from noisy transient observations.

The piezoelectric system in the acoustic regime is a coupled hyperbolic-elliptic problem. The mechanical fields obey wave equations, while the electric potential is fixed at every instant by Gauss's law, which acts as an instantaneous constraint on the mechanical strain. This makes the temporal behavior of the potential near $t=0$ matter. If the displacement grows as $\mathcal{O}(t^2)$ from rest, Gauss's law requires the potential to grow at the same order. A linear multiplier on the potential generically leaves an $\mathcal{O}(t)$ residual in the Poisson equation at early times, whereas a quadratic multiplier leaves only $\mathcal{O}(t^2)$. This compatibility between the temporal orders of the mechanical and electrical fields is referred to here as the initial-state electrostatic inconsistency, and it is examined in Sec.~\ref{sec:poisson_safeguard} and measured in Sec.~\ref{sec:ablation}.

The framework is tested in four steps. The first uses the Method of Manufactured Solutions (MMS). Because MMS supplies an exact solution, recovery of $e_{15}$ under 30~dB additive Gaussian noise can be judged without any finite-element discretization error. In this step, six training formulations are compared on identical data: two soft-penalty baselines and four constrained variants that separate what the spatial and the temporal factors each contribute (Sec.~\ref{sec:ablation}). The second step checks the full formulation against a finite-element solution of the same forced problem (Sec.~\ref{sec:fem_validation}). The third is a sweep over signal-to-noise ratios that shows how the identification degrades as the noise grows (Sec.~\ref{sec:snr_sweep}). The fourth removes the potential from the data and keeps only the displacement observations (Sec.~\ref{sec:disp_only}).

The main contributions of this work are summarized as follows:
\begin{itemize}
    \item \textbf{Controlled comparison on identical data.} Two soft-penalty baselines and four constrained variants are trained on the same five seeds, the same noisy observations, and the same initial weights. Adaptive penalties fail in four of five runs ($51.7\% \pm 31.3\%$ mean error), fixed penalties reach $0.849\% \pm 0.245\%$ only after a long second-order stage, and the constrained variants reach $0.15\%$ to $0.51\%$ with held-out $u_3$ errors below $1\%$.
    \item \textbf{What each constraint does.} The spatial factor alone gives the cleanest convergence ($0.29\%$ error after the first-order stage). The temporal factor lowers the early-time Gauss residual by a factor of 6.8 when the potential carries $t^2$ instead of $t$, but it does not change the accuracy of $e_{15}$ noticeably.
    \item \textbf{Compatibility analysis of the temporal factor.} The temporal orders of the mechanical and electrical fields near $t=0$ are analyzed for the coupled hyperbolic-elliptic system. A linear multiplier on the quasi-static potential generically leaves an $\mathcal{O}(t)$ Poisson residual, whereas a quadratic multiplier leaves an $\mathcal{O}(t^2)$ residual, and the measured residuals follow this ordering.
    \item \textbf{Non-dimensional scaling.} The scaling brings all PDE coefficients to within about one decade of $\mathcal{O}(1)$, even though the stiffnesses are of order $10^{10}$ Pa and the permittivities of order $10^{-8}$ F/m.
    \item \textbf{FEM cross-check and noise sweep.} Applied to observations on the line $\bar{x}_3 = 0.9$ from a finite-element solution of the same forced problem, with 30~dB noise, the full formulation recovers $e_{15}$ with a mean error of $1.19\% \pm 0.65\%$. A noise sweep from 40~dB to 0~dB ($N=5$ runs per level, with the initialization, collocation points, and noise redrawn each time) keeps the mean error below $1.5\%$ down to 5~dB.
    \item \textbf{Displacement-only observations.} With the potential removed from the data, $e_{15}$ is still recovered, with mean errors of $0.45\%$ on the finite-element data and $1.64\%$ on the manufactured data, although the first stage alone is then much less accurate and the recovery depends on L-BFGS.
\end{itemize}

The rest of the paper follows the same order. Section II states the governing equations and their non-dimensional form, Section III describes the constrained network, the test problems, and the training setups, Section IV presents the results, Section V states the limitations, and Section VI concludes.

\section{Theoretical Formulation}

This section defines the physical problem before any network is introduced. It states the kinematics, justifies the quasi-static approximation, writes out the coupled equations, and then rescales them, because the large gap between the mechanical and electrical coefficients comes from the PDEs themselves and not from the network.

\subsection{Infinitesimal Strain and Plane-Strain Kinematics}

\begin{figure}[htbp]
\centering
\fbox{\includegraphics[width=0.95\columnwidth]{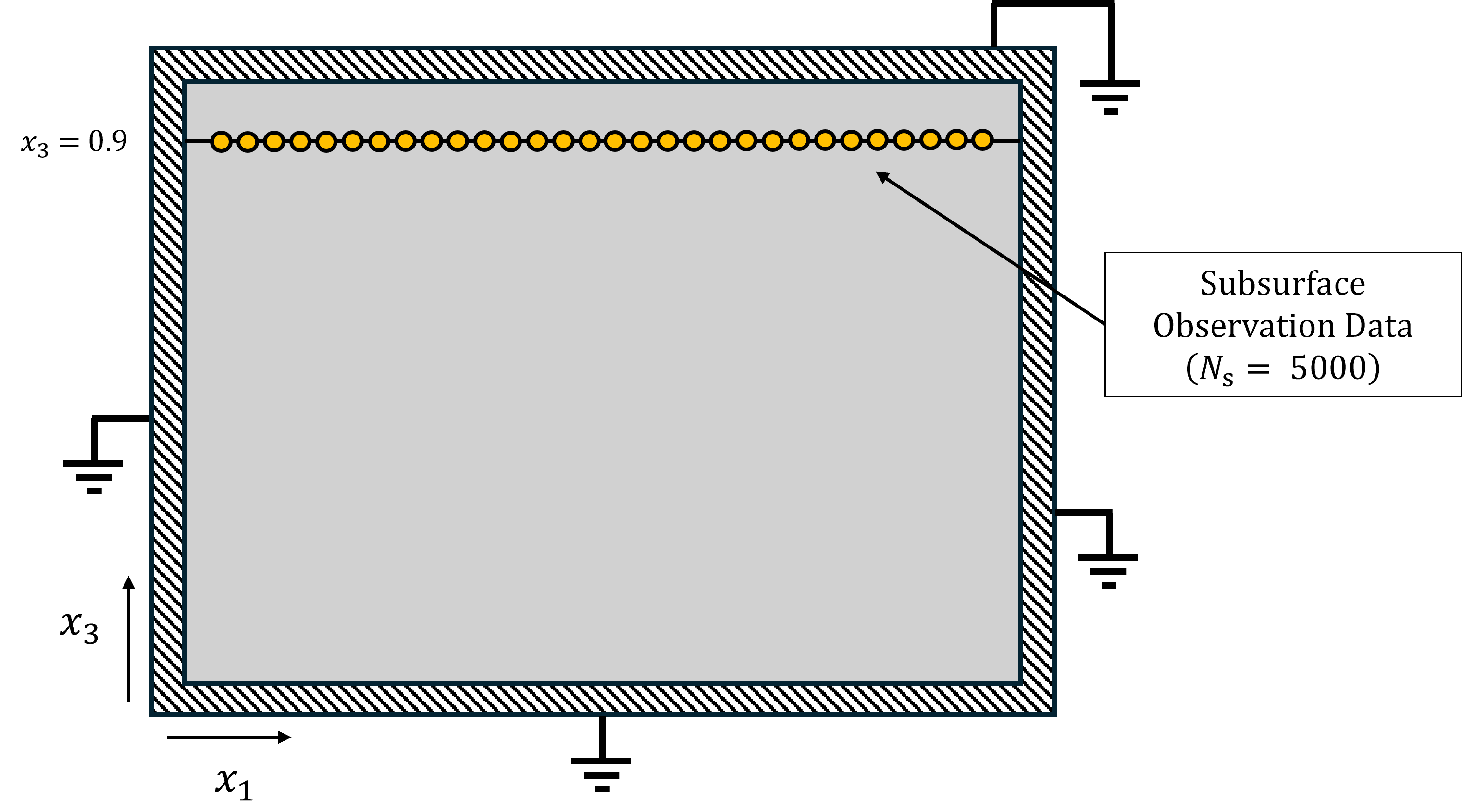}}
\caption{Schematic of the two-dimensional plane-strain computational domain for the PZT-5H medium. The perimeter is fully clamped ($u_i = 0$) and electrically grounded ($\phi = 0$), enforcing strict Dirichlet boundary conditions across the entire domain boundary. The physical orientation defines the bottom boundary at $x_3 = 0$ and the top boundary at $x_3 = 1$. For the non-destructive inverse evaluation, acoustic data are taken exclusively from a subsurface observation plane at normalized height $x_3 = 0.9$ (1 mm below the top surface of the 10 mm domain), with $N_s = 5{,}000$ space--time observation samples (not physical sensors) in the manufactured-solution tests.}
\label{fig:schematic}
\end{figure}

The transient acoustic wavefield is evaluated in a two-dimensional continuous domain $\Omega\subset\mathbb{R}^{2}$, representing the $x_{1}-x_{3}$ cross-section of a class 6mm transversely isotropic crystal (Lead Zirconate Titanate, PZT-5H), poled along the $x_{3}$-axis (see Fig.~\ref{fig:schematic}). Under the assumption of infinitesimal deformation, the Green-Lagrange strain tensor linearizes to the Cauchy strain tensor $S_{ij} = \tfrac{1}{2}(u_{i,j} + u_{j,i})$, where $\mathbf{u}(x,t)$ is the displacement vector field.

To model bulk wave mechanics while mitigating out-of-plane scattering modes, a strict state of plane strain ($u_2 = 0$, $\partial / \partial x_2 = 0$) is enforced. While SAW devices heavily exploit the out-of-plane shear-horizontal (SH) mode, the identical tensor coefficient $e_{15}$ remains fully active in the in-plane (P-SV) system through the engineering shear stress $T_5 = c_{44}(u_{1,3} + u_{3,1}) + e_{15}\phi_{,1}$ and the electric displacement $D_1$. Because $e_{15}$ is a material property, the in-plane (P-SV) system recovers the same value that governs the SH mode in later 3D device design, with a simpler and cheaper computation.

The mechanical state is completely described by the in-plane displacement vector $\mathbf{u} = (u_1, u_3)^\top$. The non-trivial strain components in Voigt notation reduce to the longitudinal strains $S_1 = u_{1,1}$ and $S_3 = u_{3,3}$, alongside the engineering shear strain $S_5 = u_{1,3} + u_{3,1}$. Plane strain leaves three unknowns ($u_1$, $u_3$, $\phi$) and keeps the 2D validation inexpensive. Going to three dimensions adds a displacement component but does not change the constrained architecture. The fully clamped, electrically grounded boundary ($u_i = 0$, $\phi = 0$ on $\partial\Omega$) is the simplest homogeneous Dirichlet setup and keeps the boundary conditions from complicating the identification of $e_{15}$. Traction-free (Neumann) boundaries, which are relevant for free-standing resonators and SAW substrates, need a different construction within the distance-function framework \cite{sukumar2022exact} and are the natural next step.

\subsection{Electrodynamic Decoupling via Asymptotic Scaling}

As the crystal lattice deforms dynamically, an internal polarization field is created. To justify using the electrostatic approximation for this dynamic system, the full electrodynamic potentials are examined. The electric field $\mathbf{E}$ is defined by the scalar potential $\phi$ and the magnetic vector potential $\mathbf{A}$ via $\mathbf{E} = -\nabla\phi - \partial\mathbf{A}/\partial t$.

The dimensionless parameter $\beta = v_a/c$ is introduced, representing the ratio of the bulk acoustic wave velocity to the speed of light in the dielectric medium. Because the acoustic wave speed is orders of magnitude slower than the speed of light in the dielectric medium, the magnetic vector potential is eliminated via standard quasi-static electrodynamic scaling \cite{tiersten1969linear, mindlin1972high}.

For piezoceramics like PZT-5H, the quasi-transverse and quasi-longitudinal acoustic wave speeds are approximately $v_t \approx 1751$~m/s and $v_L \approx 3950$~m/s. The speed of light in the dielectric medium is $c \approx c_{\mathrm{vac}}/\sqrt{\varepsilon_r} \sim 7.3 \times 10^6$~m/s. Using the upper bound $v_L$ yields a maximum ratio of $\beta \sim 5.4 \times 10^{-4} \ll 1$. Expanding the electromagnetic wave equation through asymptotic perturbation reveals that the magnetic vector potential scales as $\mathcal{O}(\beta^2)$. Taking the limit $\beta \to 0$ mathematically eliminates the transient magnetic term, reducing the electric field entirely to its irrotational component, $E_i = -\phi_{,i}$.

\subsection{The Coupled Multiphysics Boundary Value Problem}

Operating under this irrotational (quasi-static) limit, the macroscopic behavior of the continuum is governed by the linear IEEE stress-charge constitutive relations \cite{26560}. Expanding these relations for the 6mm crystal class under the derived kinematic and electrodynamic restrictions yields the following expressions for the coupled Cauchy stress ($T$) and electric displacement ($D$) tensors:

\begin{align}
T_1 &= c_{11}\,u_{1,1} + c_{13}\,u_{3,3} + e_{31}\,\phi_{,3},
  \label{eq:const_T1}\\[2pt]
T_3 &= c_{13}\,u_{1,1} + c_{33}\,u_{3,3} + e_{33}\,\phi_{,3},
  \label{eq:const_T3}\\[2pt]
T_5 &= c_{44}(u_{1,3}+u_{3,1}) + e_{15}\,\phi_{,1},
  \label{eq:const_T5}\\[2pt]
D_1 &= e_{15}(u_{1,3}+u_{3,1}) - \varepsilon_{11}\,\phi_{,1},
  \label{eq:const_D1}\\[2pt]
D_3 &= e_{31}\,u_{1,1} + e_{33}\,u_{3,3} - \varepsilon_{33}\,\phi_{,3},
  \label{eq:const_D3}
\end{align}

where $c_{ij}$ is the elastic stiffness at constant electric field, $e_{ij}$ is the piezoelectric stress constant, and $\varepsilon_{ij}$ is the dielectric permittivity at constant strain.

Substituting these constitutive laws into the conservation of linear momentum ($\rho\ddot{u}_i = T_{ij,j}$) and the charge conservation for an insulating dielectric ($D_{i,i} = 0$) yields the governing PDE system over $\Omega \times (0,T]$:

\begin{align}
\rho\ddot{u}_1 &= c_{11}\,u_{1,11} + c_{44}\,u_{1,33}
  + (c_{13}+c_{44})\,u_{3,13} \nonumber \\ 
  &\quad + (e_{31}+e_{15})\,\phi_{,13}, 
  \label{eq:pde_raw1} \\ 
\rho\ddot{u}_3 &= c_{44}\,u_{3,11} + c_{33}\,u_{3,33}
  + (c_{13}+c_{44})\,u_{1,13} \nonumber \\ 
  &\quad + e_{15}\,\phi_{,11} + e_{33}\,\phi_{,33},
  \label{eq:pde_raw3} \\ 
0 &= (e_{15}+e_{31})\,u_{1,13} + e_{15}\,u_{3,11} + e_{33}\,u_{3,33} \nonumber \\ 
  &\quad - \varepsilon_{11}\,\phi_{,11} - \varepsilon_{33}\,\phi_{,33}. 
  \label{eq:pde_raw_phi}
\end{align}

This system establishes a strictly coupled hyperbolic-elliptic differential operator. The two momentum equations are hyperbolic (wave type), while Gauss's law (\ref{eq:pde_raw_phi}) is elliptic and couples the mixed partial derivatives of the mechanical fields ($u_{1,13}$, $u_{3,11}$) to the electrostatic potential, requiring simultaneous resolution of wave propagation and the instantaneous electrostatic constraint.

\subsection{Dimensional Scaling and Stabilization}

When the original boundary value problem is turned into a training objective, the loss landscape is shaped by the sizes of the differential coefficients. For PZT-5H the stiffnesses are $\mathcal{O}(10^{10})$ Pa and the permittivities $\mathcal{O}(10^{-8})$ F/m, so the SI coefficients in the governing PDEs differ by many orders of magnitude, although they carry different units and cannot be compared directly. The mechanical and electrical residuals also carry different units, so their raw magnitudes in an unscaled mean squared error can differ by many orders of magnitude, with the exact ratio depending on the unit system and the field scales. Such imbalances are expected to make the Hessian of the objective poorly conditioned and to hinder both second-order optimizers (L-BFGS) and first-order methods (Adam).

To resolve this scale gap, the equations are non-dimensionalized. Spatial coordinates are normalized by a characteristic macroscopic length $L$ (equal to the dimensional side length of the domain), defining $x_i = L\bar{x}_i$. The temporal coordinate is normalized by the bulk quasi-transverse wave transit time $t_0 = L\sqrt{\rho/c_{44}}$ ($5.71~\mu$s for $L = 10$~mm), yielding $t = t_0\bar{t}$. The reference stiffness $c_{44}$ is deliberately selected as the normalization anchor because it directly governs the shear wave mechanics from which $e_{15}$ must be extracted. The mechanical displacement is scaled by a reference amplitude $U$ ($U = 1$~nm in all computations), defining $u_i = U\bar{u}_i$.

To enforce energetic symmetry between the hyperbolic and elliptic operators, the electric potential is scaled proportionally to the mechanical strain-induced polarization. The reference potential is defined using the primary longitudinal piezoelectric constant $e_{33}$ and the out-of-plane permittivity $\varepsilon_{33}$ as $\Phi_0 = U(e_{33}/\varepsilon_{33})$, setting $\phi = \Phi_0\bar{\phi}$. This choice ensures that the piezoelectric coupling terms in the dimensionless Poisson equation reduce to $\mathcal{O}(1)$ coefficients.

Applying the chain rule, the differential operators transform as $\partial_{x_i} = L^{-1}\partial_{\bar{x}_i}$ and $\partial_{tt} = (c_{44}/\rho L^2)\partial_{\bar{t}\bar{t}}$. Substituting these into the PDE system isolates the dimensionless normalization parameter:

\begin{equation}
k_0^2 = \frac{e_{33}^2}{c_{44}\,\varepsilon_{33}}.
\label{eq:k0}
\end{equation}

Note that $k_0^2$ is a mathematical scaling ratio defined specifically to balance the $\mathcal{O}(1)$ dimensionless Poisson equation. It should not be conflated with the standard physical electromechanical coupling coefficient $k_{33}^2$, which is strictly bounded \textit{below} unity by thermodynamic constraints. For PZT-5H, using the values $e_{33} = 23.3$ C/m$^2$, $c_{44} = 2.30\times10^{10}$ Pa, and $\varepsilon_{33}^S = 1.30\times10^{-8}$ F/m, this evaluates to $k_0^2 \approx 1.82$. With this scaling, the dimensionless PDE coefficients of Eqs.~(\ref{eq:pde_norm1})--(\ref{eq:pde_norm_phi}) span about one decade, from $\bar{e}_{15}+\bar{e}_{31} \approx 0.45$ up to $\bar{c}_{11} \approx 5.5$. This is not perfectly uniform, but it is a range in which first- and second-order optimizers are generally expected to work without per-term adaptive weighting.

The complete set of PZT-5H material constants used in this work, adapted from the IEEE piezoelectricity standard \cite{26560}, is listed in Table~\ref{tab:material}.

\begin{table}[htbp]
\centering
\caption{Material constants for PZT-5H used in the 2D plane-strain model, adapted from the IEEE piezoelectricity standard \cite{26560}, with $c_{13}$ set to the value used throughout this work (in the network, the manufactured sources, and the COMSOL model). While $c_{12}^E = 8.02\times10^{10}$~Pa is a non-zero property of the 6mm crystal class, its corresponding term identically vanishes from the governing equations under the strict plane-strain kinematic constraint ($S_2 = 0$).}
\label{tab:material}
\begin{tabular}{lcc}
\toprule
Parameter & Symbol & Value \\
\midrule
Density & $\rho$ & 7500 kg/m$^3$ \\
Stiffness (11)  & $c_{11}^E$ & $12.6\times10^{10}$ Pa \\
Stiffness (12)  & $c_{12}^E$ & $8.02\times10^{10}$ Pa \\
Stiffness (13) & $c_{13}^E$ & $5.3\times10^{10}$ Pa \\
Stiffness (33) & $c_{33}^E$ & $11.7\times10^{10}$ Pa \\
Stiffness (44) & $c_{44}^E$ & $2.30\times10^{10}$ Pa \\
Piezo. stress (15) & $e_{15}$ & $17.0$ C/m$^2$ \\
Piezo. stress (31) & $e_{31}$ & $-6.5$ C/m$^2$ \\
Piezo. stress (33) & $e_{33}$ & $23.3$ C/m$^2$ \\
Permittivity (11) & $\varepsilon_{11}^S$ & $1.503\times10^{-8}$ F/m \\
Permittivity (33) & $\varepsilon_{33}^S$ & $1.300\times10^{-8}$ F/m \\
\bottomrule
\end{tabular}
\end{table}

The finalized, dimensionally stabilized PDEs are governed entirely by dimensionless tensor ratios ($\bar{c}_{ij} = c_{ij}/c_{44}$, $\bar{e}_{ij} = e_{ij}/e_{33}$, $\bar{\varepsilon}_{ij} = \varepsilon_{ij}/\varepsilon_{33}$):

\begin{align}
\ddot{\bar{u}}_1 &= \bar{c}_{11}\,\bar{u}_{1,11} + \bar{u}_{1,33}
  + (\bar{c}_{13}+1)\,\bar{u}_{3,13} \nonumber \\
  &\quad + k_0^2(\bar{e}_{31}+\bar{e}_{15})\,\bar{\phi}_{,13},
  \label{eq:pde_norm1} \\
\ddot{\bar{u}}_3 &= \bar{u}_{3,11} + \bar{c}_{33}\,\bar{u}_{3,33}
  + (\bar{c}_{13}+1)\,\bar{u}_{1,13} \nonumber \\
  &\quad + k_0^2\bar{e}_{15}\,\bar{\phi}_{,11} + k_0^2\bar{\phi}_{,33},
  \label{eq:pde_norm3} \\
0 &= (\bar{e}_{15}+\bar{e}_{31})\,\bar{u}_{1,13}
  + \bar{e}_{15}\,\bar{u}_{3,11} + \bar{u}_{3,33} \nonumber \\
  &\quad - \bar{\varepsilon}_{11}\,\bar{\phi}_{,11} - \bar{\phi}_{,33}. 
  \label{eq:pde_norm_phi}
\end{align}

In this form every coefficient of the PDEs is $\mathcal{O}(1)$, within about one decade. This removes one leading source of ill-conditioning in the loss landscape and gives the optimizer a better-scaled gradient field, and the experiments below converge without adaptive loss weights.

The next section turns to how the boundary and initial conditions enter the network.

\section{Constrained Network, Test Problems, and Training Setup}

\subsection{Exact Boundary and Temporal Constraints}
\label{sec:poisson_safeguard}

Standard PINNs enforce boundary and initial conditions by sampling points on the space--time boundary and adding their mean squared error to the objective. In strongly coupled electro-elastodynamics this leaves the initial state only approximately enforced, which is the source of the initial-state electrostatic inconsistency discussed in Sec.~\ref{sec:intro}.

To avoid this, the initial displacement, the initial potential, and the initial rates of change of all fields are set to zero algebraically, so that Gauss's law holds trivially at $t=0$ before any gradient is evaluated. This is done by building analytical distance functions and temporal multipliers into the forward pass.

\begin{figure*}[t]
\centering
\fbox{\includegraphics[width=0.85\textwidth]{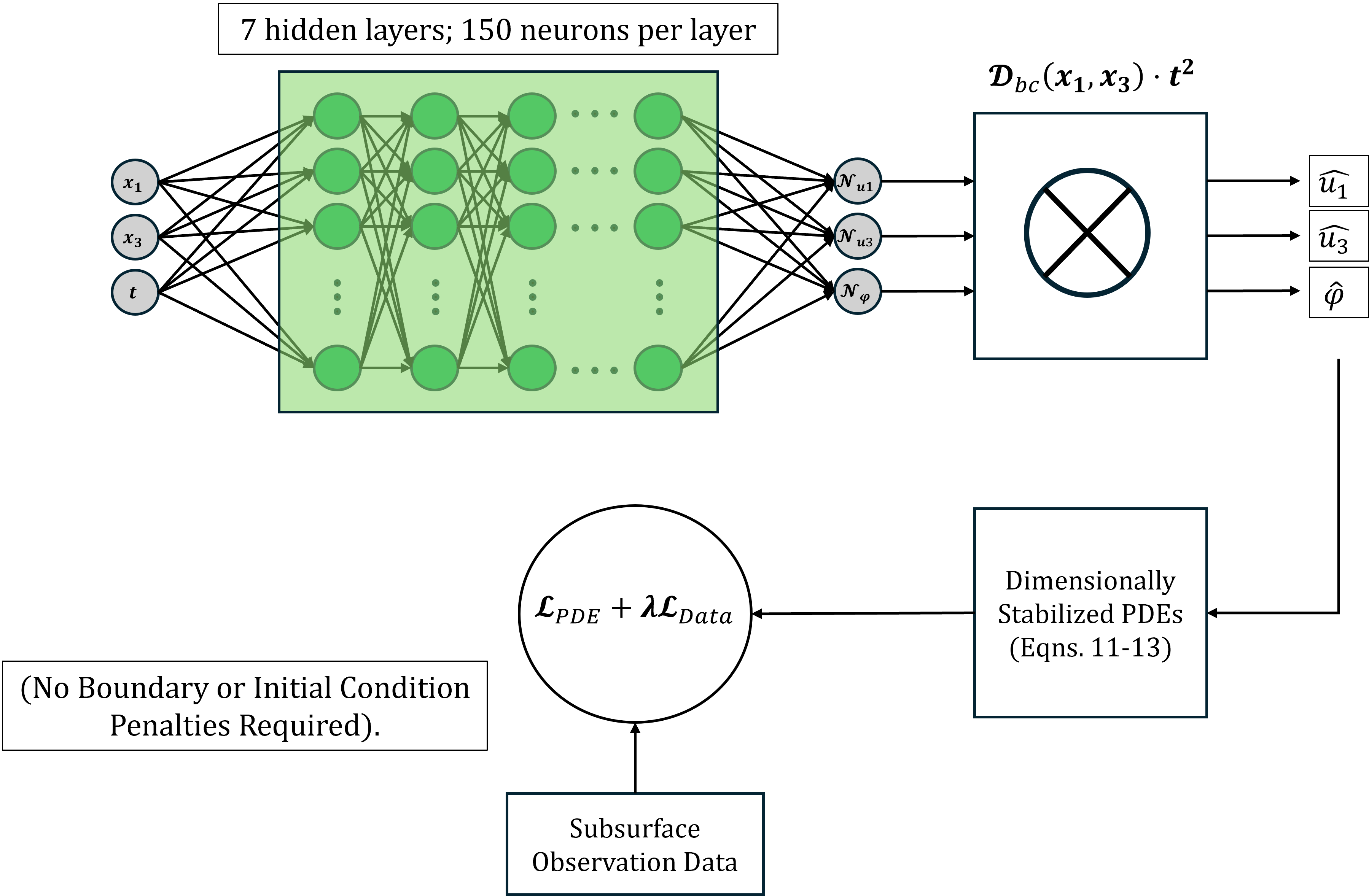}}
\caption{Topology of the constrained multiphysics neural network. The raw network outputs ($\mathcal{N}_{\theta}$) are algebraically multiplied by the exact spatio-temporal distance function ($\mathcal{D}_{bc} \cdot t^2$). This forward-pass constraint mathematically guarantees the enforcement of the clamped Dirichlet boundaries and a strictly zero-velocity initial state prior to gradient evaluation, completely removing boundary and initial conditions from the loss function.}
\label{fig:topology}
\end{figure*}

For notational simplicity in Eqs.~(\ref{eq:spatial_distance})--(\ref{eq:mms_phi}), the overbars on the normalized coordinates $(\bar{x}_1,\bar{x}_3,\bar{t})\in[0,1]^3$ and dimensionless fields are omitted where no ambiguity arises. For a normalized two-dimensional spatial domain clamped and electrically grounded at all external boundaries ($x_1 \in [0,1]$, $x_3 \in [0,1]$), the continuous spatial distance boundary multiplier is defined as:

\begin{equation}
\mathcal{D}_{\mathrm{bc}}(x_1,x_3)
  = x_1(1-x_1)\,x_3(1-x_3).
\label{eq:spatial_distance}
\end{equation}

This function is zero on the perimeter and positive inside the domain. At the center ($x_1 = x_3 = 0.5$) it equals $0.0625$ ($1/16$), so the network weights must scale up to compensate for the reduced amplitude.

Some PINN formulations use a linear temporal multiplier ($t$). The trial solution $\hat{u}_i = t \cdot \mathcal{N}_{\theta}$ then has a nonzero initial velocity, $\partial_t(\hat{u}_i)\vert_{t=0} = \mathcal{N}_{\theta} \neq 0$, and for hyperbolic wave equations an unphysical initial velocity produces a large transient residual. For the quasi-static potential, a linear multiplier already gives $\hat{\phi}(x,0)=0$ and $\nabla\hat{\phi}(x,0)=0$, so Gauss's law holds exactly at $t=0$. The issue is compatibility at early times. With $\hat{u}_i = \mathcal{O}(t^2)$, the displacement terms in Gauss's law are $\mathcal{O}(t^2)$, while a linear potential generically adds an $\mathcal{O}(t)$ term through $\varepsilon_{ij}\hat{\phi}_{,ij}$, which leaves a residual of lower order than the mechanical terms. To place all fields at the same order and to give the potential a zero initial rate of change, a second-order temporal multiplier ($t^2$) is used.

\textit{Remark on the temporal constraint applied to $\phi$.} The electric potential follows from Gauss's law, an instantaneous elliptic (quasi-static) equation that has no initial condition of its own. Since the displacement grows as $\mathcal{O}(t^2)$ near $t=0$, the induced potential does too, and so does the manufactured charge source used below, which is proportional to $[1-\cos(2\pi t)]$. The $t^2$ multiplier on $\hat{\phi}$ therefore matches the asymptotic order of the potential to that of the displacement and sets $\partial_t\hat{\phi}\vert_{t=0}=0$. It is a compatibility condition and not a physical initial condition, and the Gauss residual for $t>0$ remains a soft interior term in the loss.

The functional approximation of the unconstrained neural network is thus transformed into the exact, geometrically constrained ansatz, as illustrated in Fig.~\ref{fig:topology}.

\begin{align}
\hat{u}_1(x_1,x_3,t) &=
  \mathcal{D}_{\mathrm{bc}}(x_1,x_3)\cdot t^2
  \cdot \mathcal{N}_{u_1}(x_1,x_3,t),
  \label{eq:ansatz_u1}\\[2pt]
\hat{u}_3(x_1,x_3,t) &=
  \mathcal{D}_{\mathrm{bc}}(x_1,x_3)\cdot t^2
  \cdot \mathcal{N}_{u_3}(x_1,x_3,t),
  \label{eq:ansatz_u3}\\[2pt]
\hat{\phi}(x_1,x_3,t) &=
  \mathcal{D}_{\mathrm{bc}}(x_1,x_3)\cdot t^2
  \cdot \mathcal{N}_{\phi}(x_1,x_3,t).
  \label{eq:ansatz_phi}
\end{align}

This structure guarantees three things: (i) The spatial factor $\mathcal{D}_{\mathrm{bc}}$ enforces homogeneous Dirichlet conditions on all fields for all $t \ge 0$. (ii) At $t=0$, $\hat{u}_i = 0$ and $\hat{\phi} = 0$, so the medium starts at rest with zero potential. (iii) $\partial_t(\hat{u}_i)\vert_{t=0} = 0$ and $\partial_t(\hat{\phi})\vert_{t=0} = 0$, a zero initial rate of change for the mechanical and electrical fields alike, so Gauss's law holds at $t=0$ whatever the random initialization of the network. Because these constraints are built in, the boundary and initial conditions disappear from the loss, and the optimization acts only on the interior coupled physics.

The two factors can also be used on their own to test what each of them does. Sec.~\ref{sec:setup} defines the four constrained variants C1 to C4, and Sec.~\ref{sec:ablation} compares them.

\subsection{Manufactured-Solution Validation of Inverse Parameter Discovery}

The first test of the constrained network on the inverse problem of recovering $e_{15}$ of PZT-5H from sparse, one-sided subsurface data uses a manufactured solution, for which the exact answer is known.

To keep discretization error out of the first test, the Method of Manufactured Solutions (MMS) \cite{roache1998verification, salari2000code} is used, which prescribes an exact transient wavefield over the domain. The prescribed solution is chosen to stress the constraints: its envelope and spatial structure differ from the network's own ansatz $\mathcal{D}_{\mathrm{bc}}\,t^2\,\mathcal{N}$, and only its leading small-time behavior is proportional to $t^2$. It is a non-polynomial, spatially asymmetric wavefield with distinct spatial wavenumbers and an independent amplitude for each field component. The manufactured fields are prescribed as:

\begin{align}
u_1^*(x_1,x_3,t) &= 0.08\,[1-\cos(2\pi t)] \sin(2\pi x_1)\sin(4\pi x_3),
  \label{eq:mms_u1} \\
u_3^*(x_1,x_3,t) &= 0.10\,[1-\cos(2\pi t)] \sin(4\pi x_1)\sin(2\pi x_3),
  \label{eq:mms_u3} \\
\phi^*(x_1,x_3,t) &= 0.12\,[1-\cos(2\pi t)] \sin(2\pi x_1)\sin(2\pi x_3).
  \label{eq:mms_phi}
\end{align}

The temporal profile $[1-\cos(2\pi t)]$ is non-polynomial, but near $t=0$ it behaves as $(2\pi t)^2/2$. The network output $\mathcal{N}_\theta$ is free to represent the full sinusoidal envelope, so the $t^2$ factor does not restrict the solution to quadratic time dependence. The distinct wavenumbers and amplitudes also keep the three fields from being proportional to one another, so the cross-coupling terms that contain $e_{15}$ do not collapse onto a single spatial mode. Because $e_{15}$ enters Eqs.~(\ref{eq:pde_norm1})--(\ref{eq:pde_norm_phi}) linearly, the PDE loss is quadratic in $e_{15}$ for fixed fields and has a single minimizer whenever the sensitivity terms $k_0^2\bar{\phi}_{,13}$, $k_0^2\bar{\phi}_{,11}$, and $\bar{u}_{1,13}+\bar{u}_{3,11}$ do not vanish identically, which holds for the manufactured fields. Over $\Omega\times[0,1]$ the root-mean-square values of these three terms are $5.3$, $5.3$, and $10.4$, so $e_{15}$ is strongly constrained by the PDE residual.

Substituting these fields into the dimensionally stabilized PDEs (\ref{eq:pde_norm1})--(\ref{eq:pde_norm_phi}) via reverse-mode Automatic Differentiation generates the spatio-temporal MMS body-force terms $f_1(x_1,x_3,t)$, $f_3(x_1,x_3,t)$, and charge density $q(x_1,x_3,t)$ required to sustain the manufactured wavefield.

\begin{align}
f_1 &\equiv \ddot{\bar{u}}_1^* - \bar{c}_{11}\,\bar{u}^*_{1,11}
  - \bar{u}^*_{1,33} - (\bar{c}_{13}+1)\bar{u}^*_{3,13} \nonumber \\
  &\quad - k_0^2(\bar{e}_{31}+\bar{e}_{15}^{\mathrm{true}})\bar{\phi}^*_{,13}, 
  \label{eq:mms_f1} \\
f_3 &\equiv \ddot{\bar{u}}_3^* - \bar{u}^*_{3,11}
  - \bar{c}_{33}\,\bar{u}^*_{3,33}
  - (\bar{c}_{13}+1)\bar{u}^*_{1,13} \nonumber \\
  &\quad - k_0^2\bar{e}_{15}^{\mathrm{true}}\bar{\phi}^*_{,11} 
  - k_0^2\bar{\phi}^*_{,33},
  \label{eq:mms_f3} \\
q &\equiv -(\bar{e}_{15}^{\mathrm{true}}+\bar{e}_{31})\,\bar{u}^*_{1,13}
  - \bar{e}_{15}^{\mathrm{true}}\,\bar{u}^*_{3,11}
  - \bar{u}^*_{3,33} \nonumber \\
  &\quad + \bar{\varepsilon}_{11}\,\bar{\phi}^*_{,11} 
  + \bar{\phi}^*_{,33}.
  \label{eq:mms_q}
\end{align}

Since $\phi^*$ is prescribed independently of the mechanical fields, substituting the manufactured fields into Gauss's law gives a nonzero residual $q(x_1, x_3, t)$. In line with standard MMS practice \cite{roache1998verification, salari2000code}, $q$ is treated as a synthetic source that closes the equations. This lets the test check the optimizer and the parameter recovery on the coupled PDE operators before any finite-element effects enter, without requiring the manufactured fields to belong to the charge-free physical solution manifold. The finite-element cross-check of Sec.~\ref{sec:fem_validation} keeps the same charge source $q$ and is therefore also a charged manufactured problem.

These forcing terms are computed once from the true parameter $\bar{e}_{15}^{\mathrm{true}} = 0.72961$ and held fixed during training. The manufactured fields are evaluated over $\Omega \times [0,\bar{T}]$ with $\bar{T} = 1$, one full cycle of the $[1-\cos]$ envelope, whereas the finite-element data of Sec.~\ref{sec:fem_validation} cover $0 \le \bar{t} \le 0.876$ (0 to 5~$\mu$s).

The PINN receives these forcing terms together with a sparse set of synthetic observations $(u_1, u_3, \phi)$ collected only on the internal plane $x_3 = 0.9$. The plane is a simple access model for the subsurface interrogation motivated in Sec.~\ref{sec:intro}. The $5{,}000$ observations are space--time samples of $u_1$, $u_3$, and $\phi$ along this plane. Because the potential is fixed to zero on the whole perimeter, the interior potential is kept as an observation in the coupled data loss, and boundary electrode voltages cannot take its place.

All material constants other than $e_{15}$, including the longitudinal piezoelectric components ($e_{31}$, $e_{33}$), are treated as known, and the shear coefficient $e_{15}$ is a trainable scalar initialized at $\bar{e}_{15}^{(0)} = 0.10$ ($2.33$~C/m$^2$), an $86\%$ error in magnitude. Each tabulated result is based on five runs with different random seeds. The observations are corrupted with 30~dB additive white Gaussian noise, a moderate level used for both the MMS and the FEM tests. The signal-to-noise ratio is defined on a power basis for each field component, so 30~dB and 5~dB correspond to root-mean-square noise-to-signal ratios of about $3.2\%$ and $56\%$. 

Because the distance function enforces the Dirichlet boundaries exactly in both the MMS and the FEM tests (Sec.~\ref{sec:fem_validation}), the objective reduces to the mean squared error of the interior PDE residuals plus the data misfit at the noisy observations:

\begin{equation}
\mathcal{L}_{\mathrm{Total}}(\theta,e_{15})
  = \mathcal{L}_{\mathrm{PDE}} + \lambda\,\mathcal{L}_{\mathrm{Data}},
\label{eq:inverse_loss}
\end{equation}

where the individual loss components are defined as

\begin{align}
\mathcal{L}_{\mathrm{PDE}} &= \frac{1}{N_c}\sum_{k=1}^{N_c}
  \Bigl[\bigl(\mathcal{R}_1^{(k)}\bigr)^2
       +\bigl(\mathcal{R}_3^{(k)}\bigr)^2
       +\bigl(\mathcal{R}_\phi^{(k)}\bigr)^2\Bigr],
  \label{eq:L_PDE}\\[4pt]
\mathcal{L}_{\mathrm{Data}} &= \frac{1}{N_s}\sum_{j=1}^{N_s}
  \Bigl[\bigl(\hat{u}_1^{(j)}-u_1^{*(j)}\bigr)^2
       +\bigl(\hat{u}_3^{(j)}-u_3^{*(j)}\bigr)^2 \nonumber \\
       &\quad +\bigl(\hat{\phi}^{(j)}-\phi^{*(j)}\bigr)^2\Bigr],
  \label{eq:L_Data}
\end{align}

where $\mathcal{R}_1^{(k)}$, $\mathcal{R}_3^{(k)}$, and $\mathcal{R}_\phi^{(k)}$ are the residuals of Eqs.~(\ref{eq:pde_norm1})--(\ref{eq:pde_norm_phi}) at collocation point $k$, evaluated with the predicted fields. The momentum residuals subtract the forcing terms $f_1$ and $f_3$, and the Gauss residual adds the charge density $q$ of Eqs.~(\ref{eq:mms_f1})--(\ref{eq:mms_q}). The data weight is fixed at $\lambda = 5000$, a non-adaptive choice that gives the observations a dominant weight in the total loss. Sec.~\ref{sec:lambda} reports how the recovered parameter depends on this choice.

\subsection{FE Cross-Validation Under Additive Observation Noise and Discretization Error}
\label{sec:fem_validation}

This subsection applies the constrained network to a transient dataset computed with the COMSOL Multiphysics finite-element solver, using the same plane-strain PZT-5H geometry, clamped-and-grounded boundaries, and material constants (Table~\ref{tab:material}). 

The network is again asked to reproduce the asymmetric, non-polynomial 8-lobe wavefield of Eqs.~(\ref{eq:mms_u1})--(\ref{eq:mms_phi}). Since the manufactured fields satisfy the clamped boundaries and the zero initial conditions, they are the exact solution of the forced problem, and the finite-element output is that wavefield plus discretization error. Over all exported points and times this error is $1.97\%$ ($u_1$), $1.49\%$ ($u_3$), and $1.00\%$ ($\phi$) in relative $L^2$ norm. 

The MMS study works entirely in the normalized $\mathcal{O}(1)$ space, whereas the FEM study uses dimensional fields. The dimensionless body-force and charge-density terms ($f_1, f_3, q$) derived from Eqs.~(\ref{eq:mms_u1})--(\ref{eq:mms_phi}) are converted with the scales $c_{44}U/L^2 = 2.30\times10^{5}$~N/m$^3$ for the body forces and $e_{33}U/L^2 = 2.33\times10^{-4}$~C/m$^3$ for the charge density, with $U = 1$~nm and $L = 10$~mm, and are applied as volumetric sources in the COMSOL solid mechanics and electrostatics modules. The body forces are applied as $f_1$ and $f_3$ times the force scale, and the space charge density is applied as $-q$ times the charge scale, where $q$ is defined in Eq.~(\ref{eq:mms_q}).

The model was built in COMSOL Multiphysics 6.3 as a transient study on a 10~mm $\times$ 10~mm square, with the solid mechanics and electrostatics interfaces coupled through the piezoelectric effect and with zero initial displacement, velocity, and potential. The solution was written out every 25~ns from 0 to 5~$\mu$s. The mesh and the time-dependent solver use the physics-controlled default settings of COMSOL. The mesh has 24{,}912 triangular elements with quadratic Lagrange shape functions, which gives 150{,}675 degrees of freedom.

The fields are exported along the line $x_3 = 9$~mm ($\bar{x}_3 = 0.9$) at 5{,}000 equispaced points ($x_1 = 0.001$ to $9.999$~mm, 0.002~mm apart) and 201 equispaced times from 0 to 5~$\mu$s (normalized $\bar{T} \approx 0.876$), which gives $1.005\times10^{6}$ samples per field component. COMSOL writes the displacements in mm and the potential in volts, and both are converted to the normalized units of Sec.~\ref{sec:setup} before training.

Additive white Gaussian noise at 30~dB is added independently to each exported field component, and the noisy fields $(u_1^{\delta},\, u_3^{\delta},\, \phi^{\delta})$ are the targets of $\mathcal{L}_{\mathrm{Data}}$. As in the MMS test, the observations lie on the plane $\bar{x}_3 = 0.9$. Unlike the MMS test, which uses 5{,}000 random space--time samples, the finite-element record is dense along this line, so the data are sparse only in the sense that they come from a single plane near the top surface. Minibatches of the exported samples are drawn during training. The analytical sources $f_1$, $f_3$, and $q$ stay in the PDE residuals of Eqs.~(\ref{eq:pde_norm1})--(\ref{eq:pde_norm_phi}), so $\mathcal{L}_{\mathrm{PDE}} = 0$ at the exact manufactured solution. Any remaining PDE residual therefore comes from the discretization error of the FE data, from the noise carried into the network weights through the data loss, and from the finite capacity of the network. 

The network, the training procedure, $\lambda = 5000$, and the initial value $\bar{e}_{15}^{(0)} = 0.10$ are the same as in the MMS study (Sec.~\ref{sec:setup}).

\subsection{Computational Setup, Baselines, and Network Architecture}
\label{sec:setup}

The setup shared by all of the studies is as follows. The network is a single fully connected network with 7 hidden layers of 150 neurons and three output heads ($\hat{u}_1$, $\hat{u}_3$, $\hat{\phi}$), which gives about $137{,}000$ trainable weights in $\mathcal{N}_\theta$ (not counting the single material parameter $\bar{e}_{15}$). It uses hyperbolic tangent (Tanh) activations and Glorot uniform initialization \cite{glorot2010understanding}, and $N_c = 25{,}000$ collocation points are sampled uniformly in the interior, over $0 \le \bar{t} \le 1$ in the MMS runs and over $0 \le \bar{t} \le 0.875$ in the FEM runs (the 5~$\mu$s record ends at $\bar{t} \approx 0.876$). The non-dimensional scaling is also intended to put the gradient with respect to $\bar{e}_{15}$ on a scale comparable to that of the network weights, so that no separate learning rate is needed for $\bar{e}_{15}$ \cite{haghighat2021physics}. Once trained, the network gives the fields at any space--time query point with a single forward pass ($\sim 2.7\times10^{5}$ floating-point operations per point), independent of any mesh and without further solves.

Two soft-penalty baselines serve as the point of comparison. Both are trained on exactly the data used for the constrained network: for each seed, the same collocation points, the same $5{,}000$ noisy observations, the same noise realization, and the same initial weights. The spatiotemporal constraints are removed from the forward pass and enforced through the penalties $\mathcal{L}_{\mathrm{BC}}$ (squared fields on the four boundary faces) and $\mathcal{L}_{\mathrm{IC}}$ (squared fields and squared time derivatives at $t=0$). Baseline A adapts the weights of the mechanical and electrical PDE terms, of $\mathcal{L}_{\mathrm{BC}}$, and of the two $\mathcal{L}_{\mathrm{IC}}$ terms every 100 steps from the squared gradient norms of the individual loss terms, in the spirit of the NTK-based balancing of Wang et al. \cite{wang2022and}, with an exponential moving average of rate $0.1$, and its data weight is kept at or above $5000$. Baseline B uses no adaptation. All its penalty weights, including those of $\mathcal{L}_{\mathrm{BC}}$ and $\mathcal{L}_{\mathrm{IC}}$, are fixed at 5000, the same value as the data weight. Both baselines use the same network and the same initial guess $\bar{e}_{15}^{(0)} = 0.10$ as the constrained model, the same Adam schedule ($15{,}000$ steps, cosine decay from $10^{-3}$ to $10^{-5}$), no gradient clipping, and a full batch of $25{,}000$ collocation points, $5{,}000$ boundary points, $5{,}000$ initial-time points, and $5{,}000$ observations at every step. The L-BFGS stage that follows uses the same full batch.

The constrained variants use the same network and the same observations. They differ in which constraints are built into the forward pass. C1 is the full ansatz of Eqs.~(\ref{eq:ansatz_u1})--(\ref{eq:ansatz_phi}). C2 changes only the temporal factor of the potential from $t^2$ to $t$, with no extra penalty, so it isolates the effect of the temporal order on the electrostatics. C3 keeps the spatial factor only and adds soft penalties of weight 5000 on the initial value and the initial rate. C4 keeps the factor $t^2$ only and adds a soft boundary penalty of weight 5000. The constrained variants draw a minibatch of $5{,}000$ collocation points at each Adam step, whereas the baselines use the full batch.

\begin{figure*}[t]
    \centering
    \fbox{\includegraphics[width=0.95\textwidth]{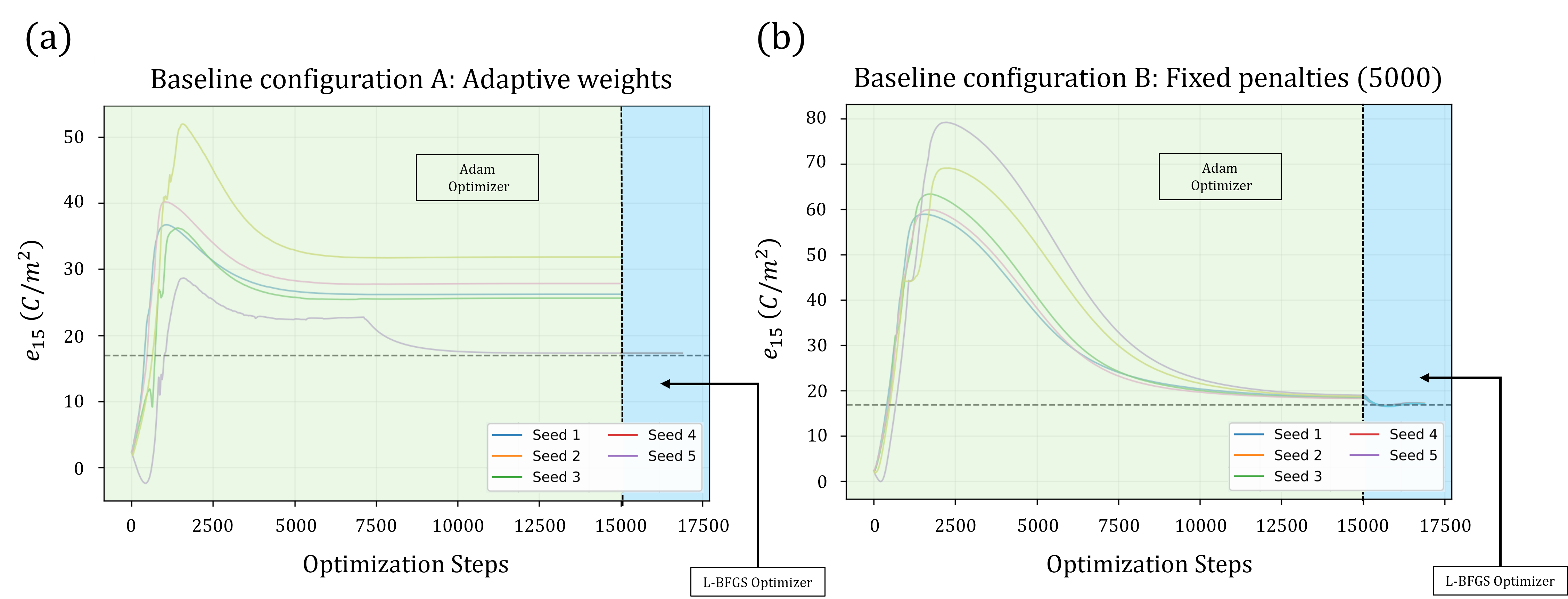}}
    \caption{Trajectories of the recovered $e_{15}$ for the two soft-penalty baselines on five seeds (30~dB noise, true value $17.0$~C/m$^2$, dashed line). The horizontal axis shows the Adam steps (0--15{,}000) followed by the L-BFGS function evaluations. (a) Baseline A with adaptive weights: four runs settle between $25.6$ and $31.9$~C/m$^2$ and one reaches $17.3$~C/m$^2$. (b) Baseline B with all penalties fixed at 5000: the parameter overshoots to between about 59 and 79~C/m$^2$, relaxes to $18.3$--$19.0$~C/m$^2$ under Adam, and ends at $17.1$--$17.2$~C/m$^2$ after L-BFGS.}
    \label{fig:baseline}
\end{figure*}

Training has two stages for every formulation. Stage 1 uses the Adam optimizer \cite{kingma2015adam} (default $\beta_1=0.9$, $\beta_2=0.999$) for 15{,}000 steps with a cosine annealing learning rate \cite{loshchilov2017sgdr} that decays from $10^{-3}$ to $10^{-5}$. In the MMS runs of the constrained network, each step uses a minibatch of 5{,}000 collocation points drawn from the 25{,}000-point pool together with all 5{,}000 observations, and in the FEM runs it uses 5{,}000 collocation points and a minibatch of 2{,}000 observation samples drawn from the exported record.

Stage 2 refines the estimate with L-BFGS and a strong-Wolfe line search \cite{nocedal2006numerical}, using the default PyTorch tolerances ($10^{-7}$ on the gradient and $10^{-9}$ on the function change) and a history size of 100. In the MMS runs of the constrained network, including arms C1 to C4, this stage works on the first 15{,}000 collocation points of the pool and all 5{,}000 observations, and in the FEM runs on 15{,}000 collocation points and a fixed random subset of 5{,}000 observation samples.

The L-BFGS stage runs for at most 1{,}500 iterations (1{,}875 function evaluations) with a learning rate of 0.1 in the MMS study and in the baseline and ablation studies, and for at most 1{,}250 iterations with a learning rate of 1 in the FEM runs of Secs.~\ref{sec:fem_comsol_results} and~\ref{sec:disp_only}.

The noise sweep (Sec.~\ref{sec:snr_sweep}), including its 30~dB entry, uses 12{,}000 Adam steps (the cosine schedule is stretched over these 12{,}000 steps) with minibatches of 4{,}000 collocation points and 2{,}000 observations, followed by at most 500 L-BFGS iterations with a learning rate of 1. All computations use float32 precision. All runs were executed on NVIDIA Tesla T4 GPUs, and the full noise sweep took about 28 hours.

To measure run-to-run variability, each formulation was trained five times. In the baseline and ablation studies of Secs.~\ref{sec:baseline} and~\ref{sec:ablation}, the five seeds fix the initial network weights, the collocation points, the observations, and the noise, and the same seeds are used for every formulation, so differences between formulations within a seed come only from the formulation. In the MMS and FEM studies of Secs.~\ref{sec:constrained_results} and~\ref{sec:fem_comsol_results}, the seed in the script was changed by hand between runs, the runs of Sec.~\ref{sec:disp_only} use the fixed seeds 1 to 5, and in the noise sweep it comes from the system clock, so in every run the initial weights, the collocation points, and the noise are redrawn. Results are given as mean $\pm$ one sample standard deviation ($N-1$ normalization) over the five runs, and the error columns report the mean of the individual absolute percentage errors.

\section{Results and Discussion}

The results follow the order of the studies described above. The two soft-penalty baselines come first, because they show what penalty enforcement does in this problem.

\subsection{Soft-Penalty Baselines}
\label{sec:baseline}

The two soft-penalty baselines of Sec.~\ref{sec:setup} are each trained on five seeds. Fig.~\ref{fig:baseline} shows their $e_{15}$ trajectories, and Table~\ref{tab:ablation} collects the final numbers.

Baseline A, with adaptive weights, ends at the wrong $e_{15}$ in four of five runs, with final errors of $54.3\%$, $50.8\%$, $63.9\%$, and $87.5\%$ (Fig.~\ref{fig:baseline}(a)). The fifth run (seed 3) reaches $1.9\%$, so the failure depends on the seed and is not guaranteed. The total loss gives no warning. It drops to about $5\times10^{-2}$ in every run, which with a data weight of 5000 means that the observations are fitted well. The fields tell a different story. At held-out space--time points, the relative $L^2$ errors are $113$--$179\%$ for $u_1$, $67$--$131\%$ for $u_3$, and $74$--$148\%$ for $\phi$, and the boundary values are off by about $0.06$--$0.10$ in dimensionless units. Even the run with a good $e_{15}$ has field errors above $100\%$, so a correct parameter does not mean correct fields. In addition, the L-BFGS stage of A stops after two to four iterations in four of the five runs and leaves $e_{15}$ unchanged.

Baseline B, with all penalties fixed at 5000, behaves differently (Fig.~\ref{fig:baseline}(b)). The parameter first overshoots to between about 59 and 79~C/m$^2$ and then relaxes, and Adam alone leaves a mean error of $9.4\% \pm 1.6\%$ (final values between $18.3$ and $19.0$~C/m$^2$). The L-BFGS stage, which runs for 1{,}123 to 1{,}214 iterations and always reaches its cap of 1{,}875 function evaluations, brings this down to $0.849\% \pm 0.245\%$ (range $0.58$--$1.24\%$). The held-out field errors are $0.9$--$2.4\%$ and the boundary values are of order $10^{-3}$. So penalty enforcement can work in this problem when the weights are chosen well and a second-order stage is available. It is the adaptive weighting of A that fails, and its result depends on the seed.

The upward bias of $e_{15}$ in A is consistent with a mechanism referred to here as a ``stiffness trap''. During early training the Dirichlet boundaries are violated, which puts phase errors into the predicted wavefield, and a larger $e_{15}$ compensates for these errors in the data loss on the observation plane, because it makes the medium stiffer and the waves faster. Piezoelectric stiffening gives the optimizer the means to do this, since the open-circuit stiffened shear modulus $c_{44}^D = c_{44}^E + e_{15}^2/\varepsilon_{11}$ grows with $e_{15}$. At the mean value of A, $e_{15} \approx 25.8$~C/m$^2$, this modulus is about 1.6 times its value at $17.0$~C/m$^2$, which corresponds to an effective shear wave speed about $26\%$ higher. The large overshoots of B point in the same direction. The experiments here do not isolate this mechanism, but the comparison in the next subsection gives some evidence, because the upward overshoot of $e_{15}$ is largest in the formulations with soft boundaries.

\subsection{Controlled Comparison of Constraint Types}
\label{sec:ablation}
\begin{figure*}[t]
    \centering
    \fbox{\includegraphics[width=0.95\textwidth]{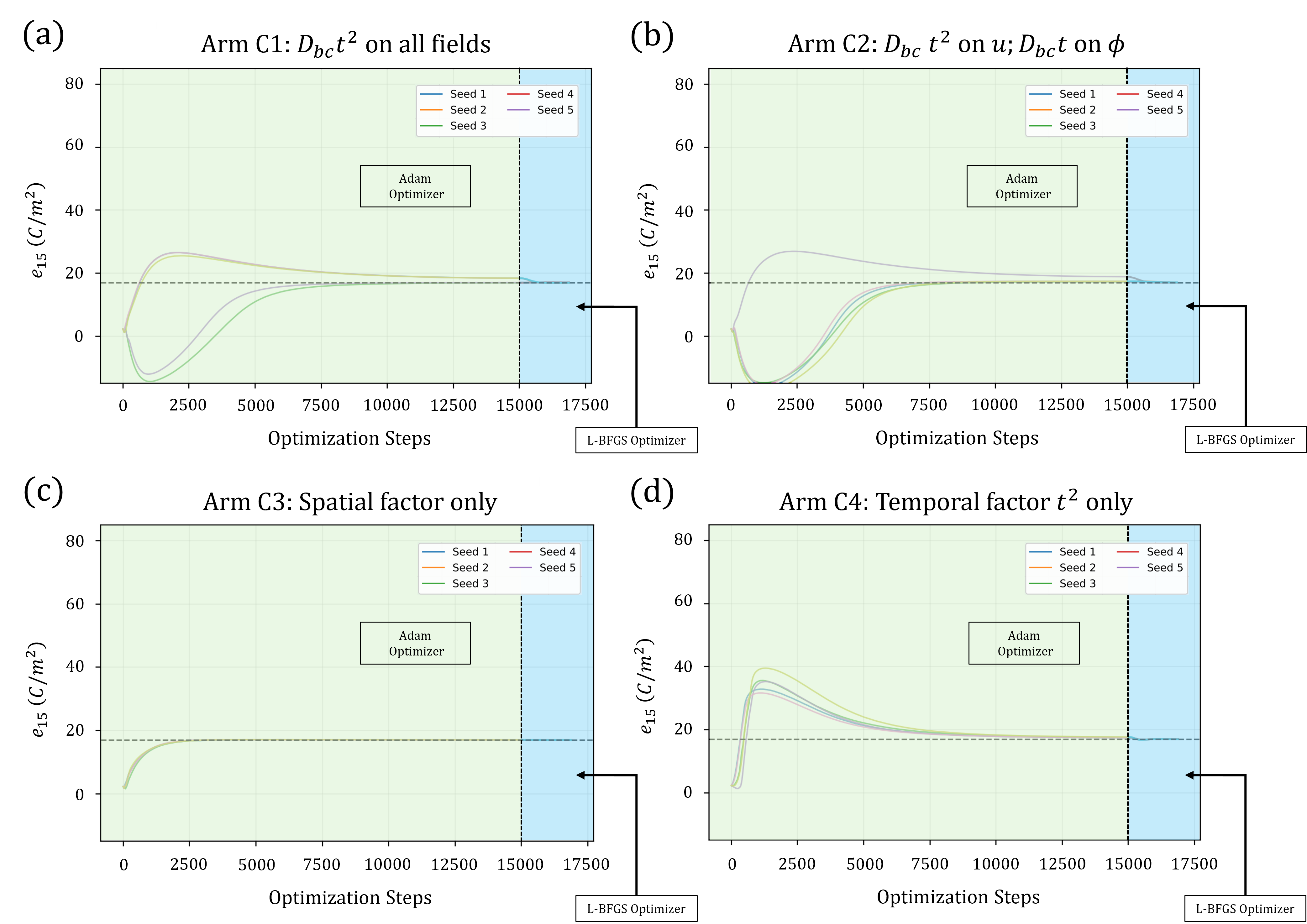}}
    \caption{Trajectories of the recovered $e_{15}$ for the four constrained variants (panels (a)--(d) show C1--C4, five seeds each, 30~dB noise). The dashed line is the true value $17.0$~C/m$^2$. The horizontal axis shows the Adam steps followed by the L-BFGS function evaluations, and the vertical axis is cut at $-15$ and $85$~C/m$^2$. C3, with the spatial factor only, converges smoothly. The variants that contain the factor $t^2$ (C1, C2, and C4) show large excursions before they settle.}
    \label{fig:ablation}
\end{figure*}

To separate what the spatial factor and the temporal factor each contribute, and to check whether the temporal order matters, the four constrained variants of Sec.~\ref{sec:setup} are trained on the same five seeds as the baselines. Table~\ref{tab:ablation} puts all six formulations side by side, and Fig.~\ref{fig:ablation} shows the $e_{15}$ trajectories of the constrained variants.

\begin{table*}[t]
\centering
\caption{Controlled comparison of six training formulations on identical data (MMS, 30~dB noise, five seeds). Errors are mean $\pm$ sample standard deviation ($N-1$) of the absolute percentage error in $e_{15}$, and ``after Adam'' is the error before the L-BFGS stage. The $u_3$ error is the mean relative $L^2$ error on 20{,}000 held-out space--time points. The boundary column is the mean root-mean-square value of $\bar{u}_3$ on the boundary faces, where the exact value is zero. The early-time residual is the root-mean-square Gauss-law residual for $t \le 0.05$, evaluated with the true manufactured sources, and is not computed for A and B. A and B are soft-penalty baselines, C1 uses $\mathcal{D}_{\mathrm{bc}}\,t^2$ on all fields, C2 uses $\mathcal{D}_{\mathrm{bc}}\,t$ on $\phi$ and $\mathcal{D}_{\mathrm{bc}}\,t^2$ on $u_1$ and $u_3$, C3 uses $\mathcal{D}_{\mathrm{bc}}$ only, and C4 uses $t^2$ only.}
\label{tab:ablation}
\begin{tabular}{llcccccc}
\toprule
Arm & Built into the network & Final error (\%) & After Adam (\%) & $u_3$ error (\%) & Boundary rms & Early-time residual \\
\midrule
A & none (adaptive penalties) & $51.7 \pm 31.3$ & $51.7 \pm 31.3$ & 85.8 & $6.9\times10^{-2}$ & -- \\
B & none (fixed penalties) & $0.849 \pm 0.245$ & $9.39 \pm 1.63$ & 1.50 & $1.1\times10^{-3}$ & -- \\
C1 & $\mathcal{D}_{\mathrm{bc}}\,t^2$ on all fields & $0.458 \pm 0.171$ & $5.14 \pm 4.40$ & 0.74 & 0 (exact) & $1.27\times10^{-2}$ \\
C2 & $\mathcal{D}_{\mathrm{bc}}\,t^2$ on $u_1,u_3$; $\mathcal{D}_{\mathrm{bc}}\,t$ on $\phi$ & $0.515 \pm 0.165$ & $3.90 \pm 4.01$ & 0.90 & 0 (exact) & $8.64\times10^{-2}$ \\
C3 & $\mathcal{D}_{\mathrm{bc}}$ only & $0.163 \pm 0.054$ & $0.285 \pm 0.115$ & 0.69 & 0 (exact) & $4.62\times10^{-2}$ \\
C4 & $t^2$ only & $0.145 \pm 0.130$ & $3.42 \pm 0.60$ & 0.61 & $5\times10^{-4}$ & $9.1\times10^{-3}$ \\
\bottomrule
\end{tabular}
\end{table*}

All four constrained variants recover $e_{15}$ to better than $0.6\%$ on average. The full formulation C1 reaches $0.458\% \pm 0.171\%$, which is close to the $0.444\% \pm 0.229\%$ of the MMS study of Sec.~\ref{sec:constrained_results}, whose five runs used different seeds. C1, C3, and C4 beat baseline B in all five seeds, and C2 does so in four of five. The fields are also more accurate. The relative $L^2$ error of $u_3$ is $0.6$--$0.9\%$ for the constrained variants and $1.5\%$ for B, and the boundary values of the variants with the spatial factor are exactly zero. The convergence behavior differs even more. After Adam alone, C3 is already at $0.29\% \pm 0.12\%$, whereas B is at $9.4\%$, so the spatial factor removes most of the need for a long second-order stage. The variants with the factor $t^2$ (C1, C2, and C4) are at $3.4$--$5.1\%$ after Adam, which is still lower than B.

The full formulation is not the best of the four, which is the less obvious result. C3 and C4, which use one exact constraint and a soft penalty for the other, reach $0.163\% \pm 0.054\%$ and $0.145\% \pm 0.130\%$, and both beat C1 in all five seeds. The trajectories in Fig.~\ref{fig:ablation} suggest where the difference comes from. In C3, $e_{15}$ rises smoothly to the true value within about 2{,}500 steps. In the three variants that contain the factor $t^2$ (C1, C2, and C4), the trajectories show large excursions. The parameter dips to roughly $-15$~C/m$^2$ or lower in some runs of C1 and C2 and overshoots in others before it settles. One plausible reason is that the factor $t^2$ makes the network output small near $t=0$, which makes the early gradients poorly scaled. This reading was not tested separately.

The temporal factor does what the compatibility analysis of Sec.~\ref{sec:poisson_safeguard} predicts. At early times ($t \le 0.05$), and with the true manufactured sources, the root-mean-square Gauss-law residual is $0.0127$ with the factor $t^2$ on the potential (C1) and $0.0864$ with the factor $t$ (C2), a ratio of 6.8. The variant without any temporal factor (C3) gives $0.046$, and the variant with $t^2$ only (C4) gives $0.0091$. This difference hardly changes $e_{15}$, though. C1 and C2 end $0.06$ percentage points apart on average ($0.458\%$ against $0.515\%$), and C2 is better than C1 in one of the five seeds.

The boundary treatment shows up in the overshoot of $e_{15}$. In the three formulations with soft boundaries (A, B, and C4) the trajectory peaks at roughly 28 to 79~C/m$^2$ before it relaxes. With the exact spatial factor the peak is at most about 27~C/m$^2$ (C1 and C2) or absent (C3). This is the direction expected from the stiffness mechanism above, but it is an observation and not a test of that mechanism.

Taken together, the comparison supports a simple conclusion. Exact constraints, whether spatial or temporal, give more accurate and more reliable recovery of $e_{15}$ than penalty enforcement, and the full formulation removes the need to tune or adapt boundary and initial penalty weights, which is where baseline A breaks down. The spatial factor gives the cleanest convergence. The temporal factor adds exact initial conditions and a small early-time electrostatic residual, at the price of a harder optimization path in these runs. The full formulation C1 is used for the finite-element cross-check and the noise sweep below, because it satisfies both the boundary and the initial conditions exactly.

Three points limit how far these numbers can be pushed. The baselines use full batches while the constrained variants use minibatches of $5{,}000$ points, so run times are not compared here. The penalty weight of 5000 in B, C3, and C4 was set equal to the data weight and was not tuned. With five runs per formulation, differences of a few hundredths of a percentage point should not be interpreted.

\subsection{Exact-Constrained Parameter Recovery}
\label{sec:constrained_results}

\begin{figure*}[t]
    \centering
    \fbox{\includegraphics[width=0.95\textwidth]{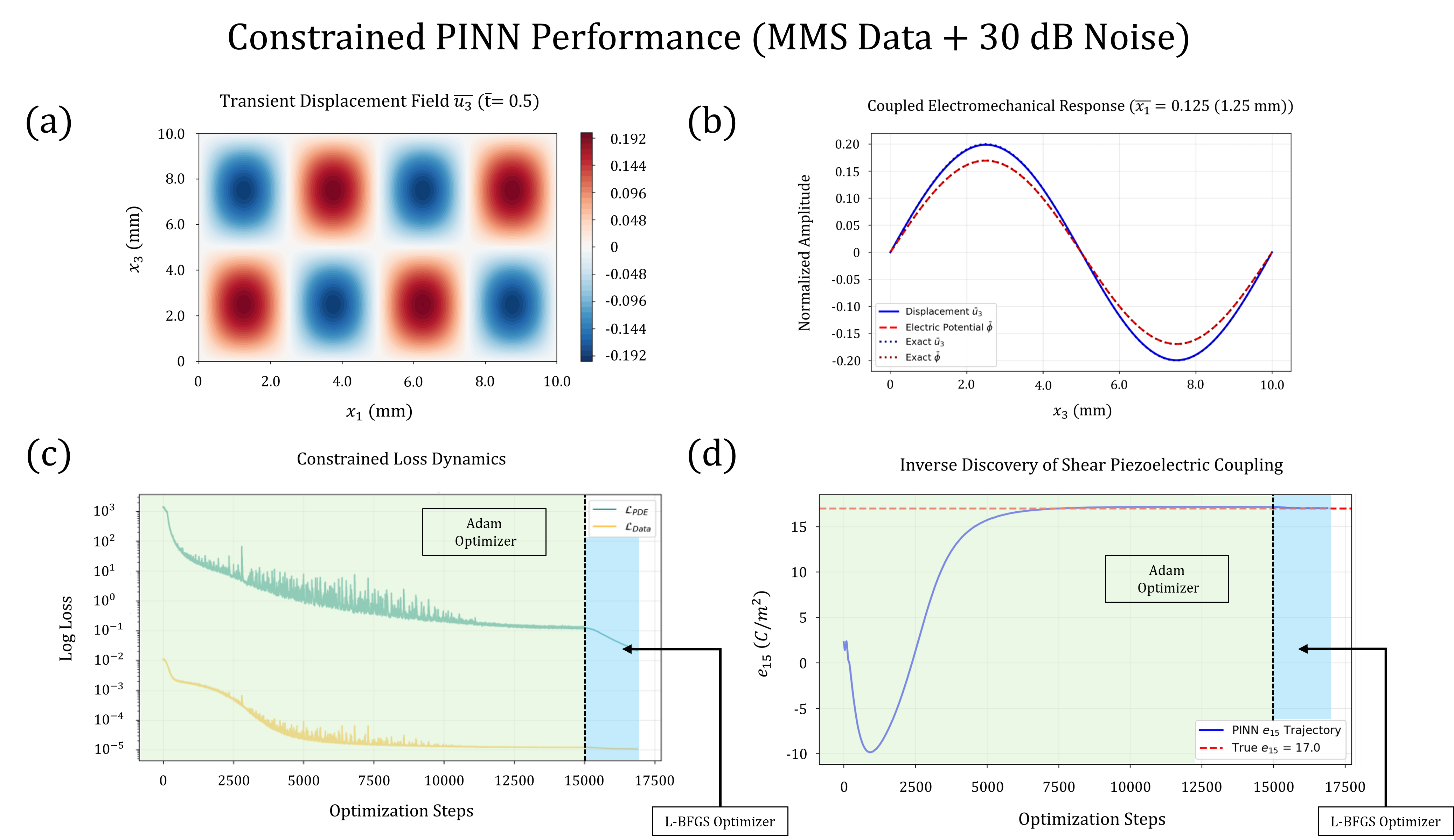}}
    \caption{Parameter discovery using the topologically constrained multiphysics PINN, evaluated against an asymmetric manufactured wavefield subject to 30~dB additive Gaussian noise. (a) The predicted dimensionless displacement field $\bar{u}_3$ at $\bar{t} = 0.5$ captures the asymmetric spatial structure arising from the independent wavenumber assignment $\sin(4\pi x_1)\sin(2\pi x_3)$, which distributes four antinodal columns across two antinodal rows, with all amplitudes exactly zero at the domain perimeter. (b) A cross-sectional profile at $\bar{x}_1 = 0.125$ ($x_1 = 1.25$~mm), where $\bar{u}_3$ has an antinode in $x_1$, shows the reconstruction of the coupled electromechanical response across the independent spatial frequencies assigned to each field. The dotted curves are the exact manufactured fields. (c) The optimization dynamics show both the internal physics loss ($\mathcal{L}_{\mathrm{PDE}}$) and the noisy observation data loss ($\mathcal{L}_{\mathrm{Data}}$) minimizing continuously through the sequential Adam and L-BFGS stages. (d) The inverse parameter trajectory for the shear coefficient $e_{15}$. The network converges to $16.975$~C/m$^2$, representing a $0.15\%$ error relative to the true target of $17.0$~C/m$^2$. The five-run mean error is $0.444\% \pm 0.229\%$.}
    \label{fig:constrained}
\end{figure*}

The constrained network removes the boundary and initial conditions from the loss, so the optimizer acts only on the interior residual and the data. Fig.~\ref{fig:constrained} shows a representative training run on the asymmetric $[1-\cos(2\pi t)]$ manufactured wavefield with 30~dB noise.

The predicted displacement $\bar{u}_3$ at $\bar{t} = 0.5$ (Fig.~\ref{fig:constrained}(a)) reproduces the asymmetric multi-lobe structure set by the wavenumber assignment $\sin(4\pi x_1)\sin(2\pi x_3)$. The multiplier $\mathcal{D}_{\mathrm{bc}}$ forces all amplitudes to zero at the four edges, so the boundary condition holds even though the interior field is strongly oscillatory and non-polynomial in time. The cross-section in Fig.~\ref{fig:constrained}(b) shows the three independently structured fields, which is consistent with the shared network representation resolving the different spatial frequencies assigned to each variable.

Both $\mathcal{L}_{\mathrm{PDE}}$ and $\mathcal{L}_{\mathrm{Data}}$ decrease throughout training (Fig.~\ref{fig:constrained}(c)). The L-BFGS stage then fine-tunes the parameter. Over five independent runs the mean error is $0.444\% \pm 0.229\%$ relative to $17.0$~C/m$^2$, and arm C1 of Sec.~\ref{sec:ablation}, which uses different seeds, gives $0.458\% \pm 0.171\%$, so the number is stable across seed sets. With the initial and boundary conditions fixed analytically, the network cannot lower the loss by introducing errors at the domain edges, so it has to satisfy the interior physics to fit the heavily weighted, noisy observations. This is consistent with the absence of the boundary-related bias seen for baseline A.

\begin{figure*}[t]
\centering
\fbox{\includegraphics[width=0.95\textwidth]{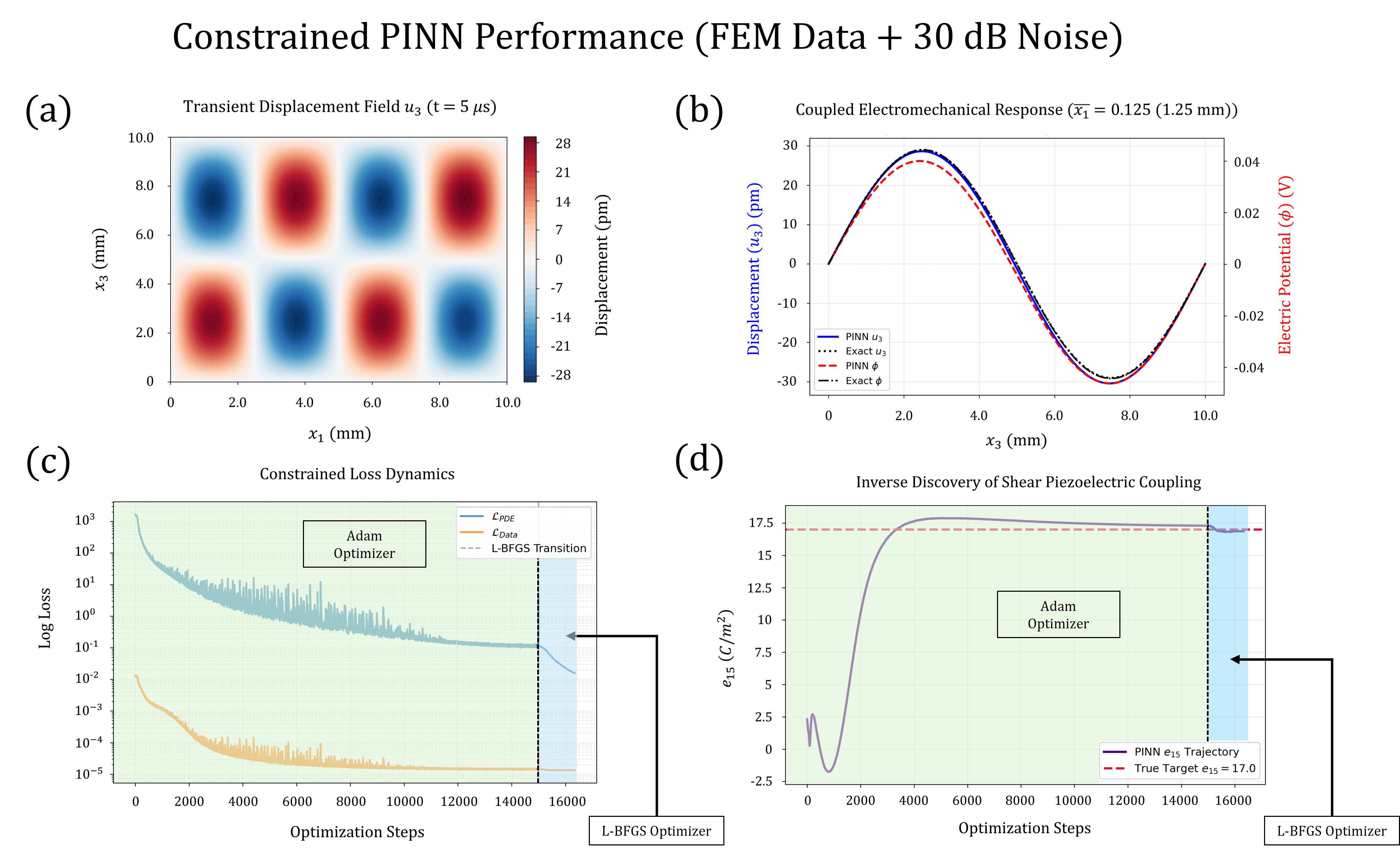}}
\caption{Parameter discovery using the identical constrained multiphysics PINN applied to COMSOL FEM transient data of the same forced problem with 30~dB additive Gaussian noise. (a) The predicted displacement field $u_3$ at $t = 5.0\;\mu\mathrm{s}$ (normalized $\bar{t} \approx 0.876$) reproduces the asymmetric 8-lobe standing wave pattern across the domain and vanishes at the clamped boundaries, as enforced by $\mathcal{D}_{\mathrm{bc}}\cdot t^2$. (b) A cross-sectional profile evaluated explicitly at $x_1 = 1.25$~mm (the primary spatial antinode of the $u_3$ wavefield) shows co-phase behavior of the displacement and the potential. The dotted curves are the exact manufactured fields. The dual axes show the physical amplitudes ($u_3$ in pm, $\phi$ in V). (c) Both $\mathcal{L}_{\mathrm{PDE}}$ and $\mathcal{L}_{\mathrm{Data}}$ decrease through the Adam phase, apart from minibatch noise, and stabilize under L-BFGS. (d) The inverse parameter trajectory for the shear coefficient $e_{15}$. The network converges from an initial estimate of $2.33$~C/m$^2$ to $16.873$~C/m$^2$, representing a $0.75\%$ error relative to the true target of $17.0$~C/m$^2$.}
\label{fig:fem_comsol}
\end{figure*}

\subsection{Data-Weighting Sensitivity}
\label{sec:lambda}

The data weight $\lambda$ is the one weight set by hand in the loss of the full formulation, so this subsection checks how the recovered parameter depends on it. The test uses the noisy asymmetric manufactured wavefield at 30~dB.

Table~\ref{tab:lambda} lists the recovered $e_{15}$ over a decade of $\lambda$ values. The mean error falls monotonically with $\lambda$ over the tested range, from $1.497\%$ at $\lambda=1000$ to $0.175\%$ at $\lambda=10000$, and every value stays within $1.5\%$ of the true parameter. The value $\lambda = 5000$ is an intermediate setting inside this range and was kept for the FEM cross-check and the SNR sweep. A larger weight would lower the error further at this noise level, but 5000 was kept so that the other studies stay comparable.

\begin{table}[htbp]
\centering
\caption{Sensitivity of the recovered $e_{15}$ to the data-weighting hyperparameter $\lambda$. Results are reported as mean~$\pm\,1\sigma$ over five independent training runs evaluated on the asymmetric MMS under 30~dB additive Gaussian noise. The $\lambda = 5000$ row is the five-run MMS study of Sec.~\ref{sec:constrained_results}. The true dimensional reference value is $e_{15}^{\mathrm{true}} = 17.0$~C/m$^2$.}
\label{tab:lambda}
\begin{tabular}{ccc}
\toprule
$\lambda$ & Mean $e_{15}$ & Mean Absolute Error \\
 & (C/m$^2$) & (\%) $\pm\,1\sigma$ \\
\midrule
1000 & 17.254 & $1.497 \pm 0.750$ \\
2500 & 17.039 & $0.952 \pm 0.219$ \\
5000 & 16.981 & $0.444 \pm 0.229$ \\
10000 & 16.970 & $0.175 \pm 0.045$ \\
\bottomrule
\end{tabular}
\end{table}

\subsection{FEM Cross-Check at 30~dB Noise}
\label{sec:fem_comsol_results}

Fig.~\ref{fig:fem_comsol} shows a representative run of the FEM cross-check. The network reproduces the asymmetric 8-lobe transient displacement field (Fig.~\ref{fig:fem_comsol}(a)). The data come from a finite-element computation, yet the displacement goes cleanly to zero at all four edges, as expected: the $\mathcal{D}_{\mathrm{bc}}$ multiplier applies the exact boundary conditions, which the manufactured solution satisfies.

\begin{figure*}[t]
\centering
\fbox{\includegraphics[width=0.95\textwidth]{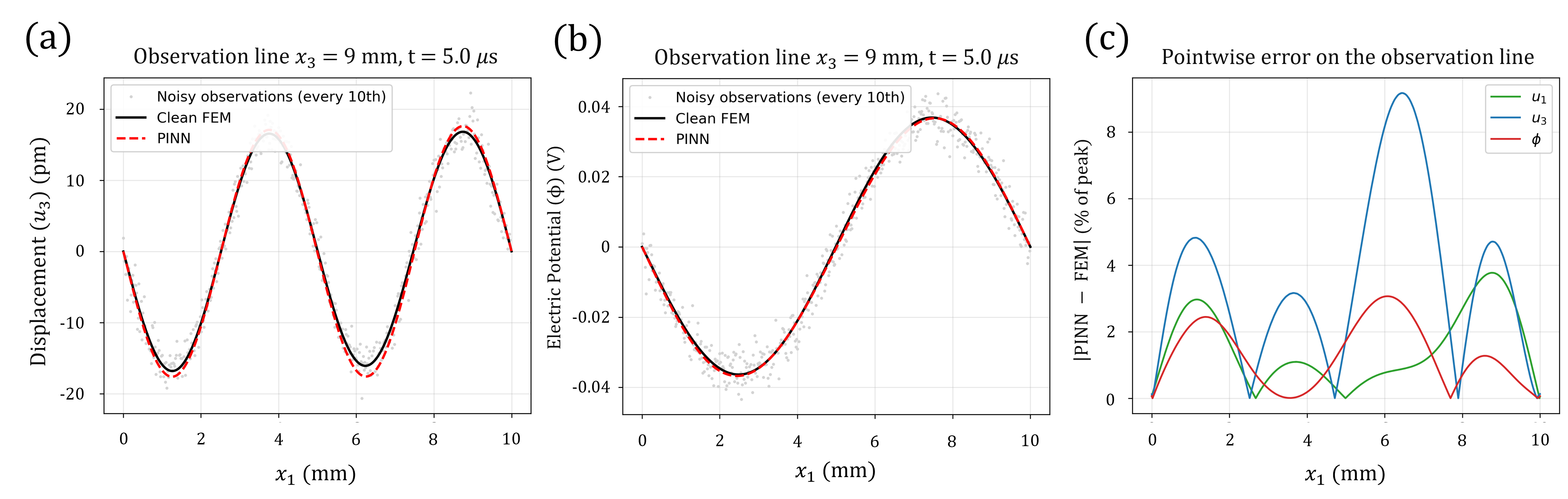}}
\caption{Fields along the observation line $x_3 = 9$~mm at $t = 5.0\;\mu\mathrm{s}$ for the representative finite-element run. (a), (b) Clean FEM field, every tenth noisy observation, and the PINN prediction for $u_3$ and $\phi$. (c) Pointwise difference between the PINN and the clean FEM field, as a percentage of the peak value of each field.}
\label{fig:fem_line}
\end{figure*}

The cross-section in Fig.~\ref{fig:fem_comsol}(b), taken at the spatial antinode $x_1=1.25$~mm to avoid nodal lines, shows the displacement and the potential in phase. The dotted curves are the exact manufactured fields at the same position and time. At $t=5.0\;\mu$s the wavefield is small, and the largest difference between the PINN and the exact curve is $5.5\%$ of the peak for $u_3$ and $11.4\%$ of the peak for $\phi$. The potential is therefore reproduced with a larger relative error than the displacement at this time. A single late snapshot also gives larger relative errors than the relative $L^2$ errors over the whole record reported below, because the amplitude is small and part of the difference comes from the finite-element data, which differ from the exact field by $1$--$2\%$ in relative $L^2$ norm.

Both $\mathcal{L}_{\mathrm{PDE}}$ and $\mathcal{L}_{\mathrm{Data}}$ decrease steadily apart from the minibatch noise of Adam (Fig.~\ref{fig:fem_comsol}(c)). The PDE loss levels off at a nonzero floor, as expected when continuous analytical source terms are reconciled with finite-element target data. 

In the representative run of Fig.~\ref{fig:fem_comsol}(d), $e_{15}$ converges rapidly during the Adam phase and L-BFGS refines it. The recovered value stays accurate despite the noise that remains in the displacement field.

Fig.~\ref{fig:fem_line} separates the clean finite-element field, the noisy observations, and the PINN prediction along the observation line $x_3=9$~mm at $t=5.0\;\mu$s. At this time the added noise has a root-mean-square value of about $10\%$ of the peak of each field, because the 30~dB level is set by the signal power of the whole record. The PINN deviates from the clean field by $1.6$--$4.3\%$ of the peak in root mean square, and by at most $9.2\%$ for $u_3$ (Fig.~\ref{fig:fem_line}(c)). Over all 5{,}000 points and 201 times of the record, the relative $L^2$ errors against the clean finite-element data are $1.01\%$, $1.17\%$, and $0.87\%$ for $u_1$, $u_3$, and $\phi$ in this run.

Table~\ref{tab:fem_comparison} compares the MMS and FEM runs of the full formulation. The first FEM row is the 30~dB entry of the noise sweep (Sec.~\ref{sec:snr_sweep}), so it comes from the shorter sweep schedule of 12{,}000 Adam steps, whereas the representative run in Fig.~\ref{fig:fem_comsol} uses the 15{,}000-step schedule of Sec.~\ref{sec:setup}. The MMS study stays below $0.5\%$. The error rises to $1.189\% \pm 0.651\%$ for the finite-element data at the same noise level. The finite-element fields differ from the exact manufactured fields by $1.0$--$2.0\%$ in relative $L^2$ norm, of the same order as the $3.2\%$ noise amplitude at 30~dB, and the two tests also differ in observation set and time window, so the increase reflects all of these factors together. In the MMS and FEM studies, each run redraws the network initialization, the collocation points, and the noise, so the standard deviation is the combined run-to-run variability.

\begin{table}[htbp]
\centering
\caption{Inverse $e_{15}$ identification with the full constrained formulation on the manufactured (MMS) and finite-element (FEM) data. Results are reported as mean~$\pm\,1\sigma$ over $N=5$ independent training runs evaluated at 30~dB additive Gaussian noise. The error column reports the mean of individual absolute percentage errors $\langle |e_{15,i} - e_{15}^{\mathrm{true}}| / e_{15}^{\mathrm{true}} \rangle$ across the $N=5$ runs. The true dimensional reference value is $e_{15}^{\mathrm{true}} = 17.0$~C/m$^2$.}
\label{tab:fem_comparison}
\begin{tabular}{lcc}
\toprule
Dataset & Mean $e_{15}$ & Mean Error (\%) \\
 & (C/m$^2$) & $\pm\,1\sigma$ \\
\midrule
MMS & $16.981$ & $0.444 \pm 0.229$ \\
FEM, sweep schedule & $16.810$ & $1.189 \pm 0.651$ \\
FEM, long schedule & $16.837$ & $0.959 \pm 0.309$ \\
\bottomrule
\end{tabular}
\end{table}

The FEM cross-check used a single noise level, and Sec.~\ref{sec:snr_sweep} varies this noise level.

\subsection{Noise Sensitivity Across the Full SNR Spectrum}
\label{sec:snr_sweep}

\begin{figure*}[htbp]
\centering
\fbox{\includegraphics[width=0.8\linewidth]{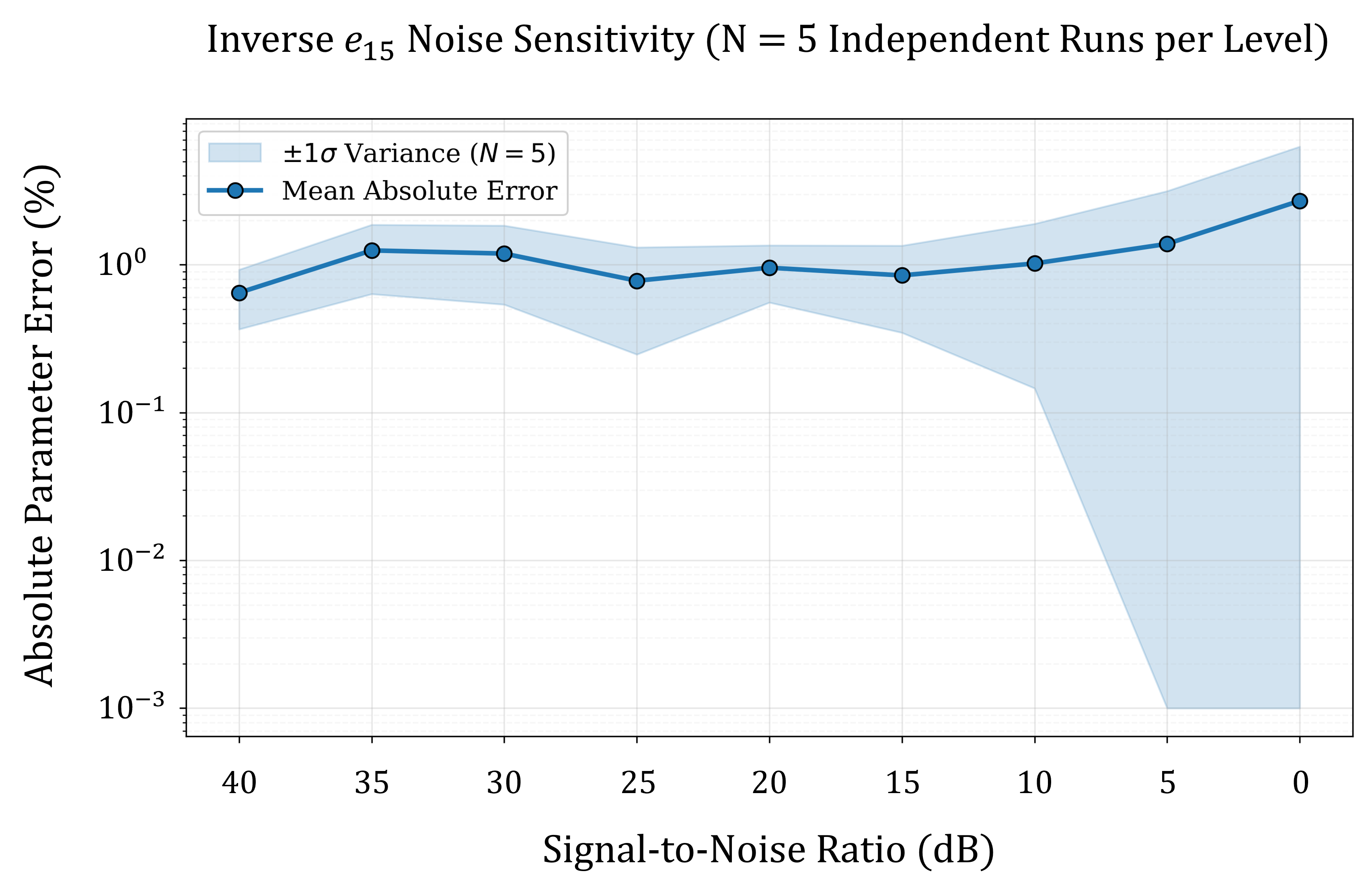}}
\caption{Statistical noise sensitivity of the constrained PINN inverse $e_{15}$ discovery across nine SNR levels from 40~dB down to 0~dB. The y-axis displays the absolute percentage error in the recovered $e_{15}$ on a logarithmic scale, and the x-axis is inverted to reflect increasing noise. The solid line represents the mean absolute error evaluated over $N=5$ training runs per noise level, while the shaded region denotes the $\pm 1\sigma$ band, which is truncated at the lower axis limit where the mean minus one standard deviation is not positive.}
\label{fig:snr_sweep}
\end{figure*}

To see how the identification depends on noise, the same constrained network, training pipeline, and finite-element wavefield are applied at nine SNR levels from 40~dB down to 0~dB. Before each training run a fresh, independent noise realization is added to all three field components. Table~\ref{tab:snr_sweep} lists the recovered $e_{15}$ and its absolute error at each level, and Fig.~\ref{fig:snr_sweep} shows the error curve on a logarithmic scale.

\begin{table}[htbp]
\centering
\caption{Statistical noise sensitivity of the constrained PINN inverse $e_{15}$ discovery across nine SNR levels. Results are reported as mean~$\pm\,1\sigma$ over $N=5$ independent training runs per noise level, each evaluating a completely reinitialized network against a freshly sampled Gaussian noise realization. The error column reports the mean of individual absolute percentage errors $\langle |e_{15,i} - e_{15}^{\mathrm{true}}| / e_{15}^{\mathrm{true}} \rangle$ across the $N=5$ runs. The true reference value is $e_{15}^{\mathrm{true}} = 17.0$~C/m$^2$.}
\label{tab:snr_sweep}
\begin{tabular}{ccc}
\toprule
SNR (dB) & Mean $e_{15}$ (C/m$^2$) & Error (\%) $\pm\,1\sigma$ \\
\midrule
40 & 16.973 & $0.646 \pm 0.280$ \\
35 & 16.838 & $1.249 \pm 0.615$ \\
30 & 16.810 & $1.189 \pm 0.651$ \\
25 & 16.952 & $0.779 \pm 0.532$ \\
20 & 16.861 & $0.954 \pm 0.399$ \\
15 & 16.856 & $0.847 \pm 0.500$ \\
10 & 17.021 & $1.019 \pm 0.873$ \\
\phantom{0}5  & 16.870 & $1.383 \pm 1.768$ \\
\phantom{0}0  & 17.237 & $2.705 \pm 3.594$ \\
\bottomrule
\end{tabular}
\end{table}

With $N=5$ runs per level, the spread shows how much the result varies from run to run at each noise level. Since the initialization, the collocation points, and the noise are all redrawn in every run, the spread is the combined variability. As Fig.~\ref{fig:snr_sweep} and Table~\ref{tab:snr_sweep} show, the mean error does not rise monotonically between 40~dB and 10~dB and stays in overlapping bands of about $0.6\%$ to $1.2\%$. This plateau is consistent with an accuracy floor set by the $1.0$--$2.0\%$ discretization error of the finite-element data. This explanation was not tested separately.

At lower SNR the noise energy approaches the signal energy and the mean error drifts upward, to $2.71\%$ at 0~dB. The spread grows markedly too: at 5~dB the standard deviation ($1.768\%$) exceeds the mean ($1.383\%$), and at 0~dB it is $3.594\%$. The run errors are therefore heavy-tailed, and the means should not be read as bounds on individual runs. Even so, the mean error stays at the percent level down to equal signal and noise power.

\subsection{Displacement-Only Observations and Field Accuracy}
\label{sec:disp_only}

\begin{figure*}[t]
\centering
\fbox{\includegraphics[width=0.95\textwidth]{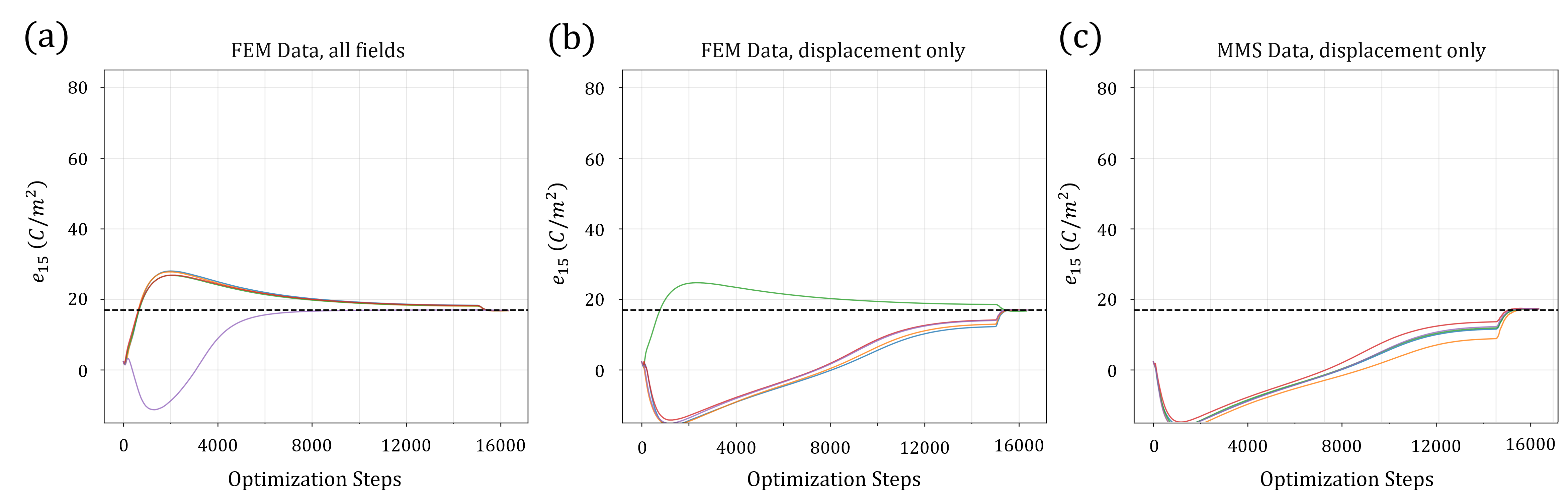}}
\caption{Trajectories of the recovered $e_{15}$ for five seeds at 30~dB noise. (a) FEM data with all three fields observed. (b) FEM data with displacement observations only. (c) MMS data with displacement observations only. The horizontal axis shows the Adam steps followed by the L-BFGS function evaluations, and the Adam stage ends at step 15{,}000. The dashed line is the true value of $17.0$~C/m$^2$, and the vertical axis is cut at $-15$ and $85$~C/m$^2$.}
\label{fig:disp_only}
\end{figure*}

The electric potential cannot be measured inside a real device (Sec.~\ref{sec:limitations}), so the full formulation was also trained on displacement observations only. The data term $\mathcal{L}_{\mathrm{Data}}$ then contains only $u_1$ and $u_3$. Everything else is unchanged: the PDE residuals of all three equations, the exact constraints, $\lambda = 5000$, and the schedules. The runs use seeds 1 to 5 at 30~dB on the manufactured data and on the finite-element data. On the manufactured data the seeds, collocation points, and noise are those of arm C1 in Sec.~\ref{sec:ablation}, so C1 is the reference with all three fields observed. On the finite-element data, the reference is a new set of runs with all three fields and the same seeds, which uses the schedule of the representative run (15{,}000 Adam steps and at most 1{,}250 L-BFGS iterations). Table~\ref{tab:disp_only} lists the results and Fig.~\ref{fig:disp_only} shows the trajectories.

\begin{table*}[t]
\centering
\caption{Recovery of $e_{15}$ with and without the potential in the data (30~dB, seeds 1 to 5). Errors are mean $\pm$ sample standard deviation ($N-1$) of the absolute percentage error in $e_{15}$, and ``after Adam'' is the error before the L-BFGS stage. The last three columns are the mean relative $L^2$ errors of the fields (\%) against the exact wavefield at 20{,}000 held-out space--time points. The MMS rows with all fields are arm C1 of Table~\ref{tab:ablation}. The FEM rows use the schedule of the representative run.}
\label{tab:disp_only}
\begin{tabular}{llccccc}
\toprule
Data & Observed fields & Final error (\%) & After Adam (\%) & $u_1$ & $u_3$ & $\phi$ \\
\midrule
MMS & $u_1,u_3,\phi$ & $0.458 \pm 0.171$ & $5.14 \pm 4.40$ & 0.99 & 0.74 & 0.71 \\
MMS & $u_1,u_3$ & $1.64 \pm 0.30$ & $31.5 \pm 10.3$ & 1.57 & 1.15 & 1.43 \\
FEM & $u_1,u_3,\phi$ & $0.959 \pm 0.309$ & $5.68 \pm 3.20$ & 0.96 & 0.88 & 0.64 \\
FEM & $u_1,u_3$ & $0.452 \pm 0.727$ & $19.0 \pm 7.1$ & 1.00 & 0.88 & 0.89 \\
\bottomrule
\end{tabular}
\end{table*}

Without the potential, $e_{15}$ is still recovered. On the finite-element data the mean error is $0.452\% \pm 0.727\%$, against $0.959\% \pm 0.309\%$ with all fields. The spread is wider in the displacement-only case because of seed 3 ($1.75\%$), while the other four seeds give $0.05$--$0.21\%$. With five seeds the two means should not be ranked. On the manufactured data the displacement-only error is $1.64\% \pm 0.30\%$, larger than the $0.458\% \pm 0.171\%$ of C1, and larger in each of the five seeds ($1.29$--$2.01\%$ against $0.26$--$0.69\%$). The potential data therefore help in the manufactured test, and their effect on the finite-element data is not clear at this sample size.

The first stage behaves differently without the potential. After Adam alone the error is $19.0\% \pm 7.1\%$ on the finite-element data and $31.5\% \pm 10.3\%$ on the manufactured data, against $5.7\%$ and $5.1\%$ with all fields, and in most runs the trajectory is still below the true value when Adam ends (Fig.~\ref{fig:disp_only}(b), (c)). The L-BFGS stage then completes the recovery, so the displacement-only identification depends on the second-order stage, as baseline B does. The held-out field errors change little (Table~\ref{tab:disp_only}). The potential, which is no longer observed, is still reproduced to $0.9\%$ on the finite-element data and $1.4\%$ on the manufactured data, so the equations carry it from the displacement.

For the field accuracy with all three fields observed on the finite-element data, the relative $L^2$ errors against the exact wavefield at held-out points are $0.96\%$, $0.88\%$, and $0.64\%$ for $u_1$, $u_3$, and $\phi$. These are smaller than the differences between the finite-element solution and the exact field ($1.97\%$, $1.49\%$, and $1.00\%$), which is consistent with the PDE residuals being evaluated with the exact forcing, so that the network is pulled toward the continuous solution and not toward the discretization error. This was not tested separately. Against the noise-free finite-element export on the observation line, the errors are $1.07\%$, $1.21\%$, and $0.88\%$. Both sets of numbers are means over seeds 1 to 5. Their mean $e_{15}$ error of $0.959\% \pm 0.309\%$ is the second FEM row of Table~\ref{tab:fem_comparison}, and it differs from the $1.189\% \pm 0.651\%$ of the first row because the schedule and the seeds differ.

\section{Limitations and Outlook}
\label{sec:limitations}

The study rests on several choices that limit what it shows. This section states each one and the next step it suggests.

\subsection{Boundaries and Excitation}

The formulation handles only homogeneous Dirichlet boundaries, a rest initial state, and a volumetric source. Real excitations such as electrode voltage pulses, surface tractions, or laser-generated loading act through the boundaries. The next step is to extend the exact construction to non-homogeneous boundary data and, for free surfaces, to Neumann conditions.

\subsection{Observations and Data}

In the main runs the observations include the interior electric potential, which cannot be measured inside a real device. The displacement-only test of Sec.~\ref{sec:disp_only} removes it and $e_{15}$ is still recovered, but it uses the same simulated data and noise model. Because the perimeter potential is fixed to zero here, boundary electrode voltages cannot stand in for it, so a measurement model built on quantities that can be recorded is needed. The finite-element data also solve the same forced problem as the manufactured solution, so that test checks the solver and not a different excitation, and no experimental data are used.

\subsection{Scope of the Comparison}

The comparison of formulations uses one noise level, one manufactured field, and five seeds per formulation, and the soft penalty weights were set to 5000 without further tuning. Repeating it with more seeds, other fields, and tuned weights would show how general the ranking is. The finite-element cross-check and the noise sweep use only the full formulation, and running them for the spatial-only variant is the obvious next test.

\subsection{Material Constants and Geometry}

The constants $c_{44}$, $e_{31}$, $e_{33}$, and the dielectric constants are treated as known, so a sensitivity study of how errors in these constants carry over to $e_{15}$ is also needed. The test geometry is a millimeter-scale block of bulk PZT-5H, and micrometer-scale PMUT stacks and traction-free SAW substrates still have to be tried.

\section{Conclusion}

This paper presents a constrained multiphysics PINN for identifying the shear piezoelectric coefficient $e_{15}$ from sparse, noisy wavefield observations, and compares it with penalty-based alternatives on identical data. A soft-penalty baseline with adaptive loss weights failed in four of five runs ($51.7\% \pm 31.3\%$ mean error) and returned fields with relative $L^2$ errors above $65\%$ at held-out points. With fixed penalties, the baseline reached $0.849\% \pm 0.245\%$, but only after a long second-order stage, since the first-order stage alone left $9.4\%$ error. The constrained variants reached mean errors between $0.15\%$ and $0.51\%$ with held-out $u_3$ errors below $1\%$. 

The constrained formulation has three parts. First, exact spatial Dirichlet enforcement through boundary-conforming distance functions \cite{sukumar2022exact, lagaris1998artificial}, together with hard enforcement of initial conditions as shown for elastodynamics \cite{rao2021physics} and in the author's earlier one-dimensional electro-elastodynamic work \cite{bharadwaj2026thermodynamically}, is extended to a two-dimensional coupled piezoelectric system. Second, a second-order temporal multiplier ($t^2$) places the displacement and the quasi-static potential at the same asymptotic order near $t=0$ and sets their initial rates of change to zero. Third, a non-dimensional scaling brings the PDE coefficients to within about one decade of $\mathcal{O}(1)$. The controlled comparison adds two findings about these parts. The spatial factor alone gives the cleanest convergence, and the temporal factor lowers the early-time electrostatic residual by a factor of about seven relative to a linear factor without noticeably changing the accuracy of $e_{15}$.

On a PZT-5H plane-strain system with a spatially asymmetric manufactured wavefield and 30~dB Gaussian noise, the full formulation recovers $e_{15}$ from $5{,}000$ space--time observations on a subsurface plane with a mean error of $0.444\% \pm 0.229\%$, and arm C1 of the controlled comparison, with different seeds, gives $0.458\% \pm 0.171\%$. On a finite-element solution of the same forced problem at the same noise level, the mean error is $1.19\% \pm 0.65\%$. When only the displacement is observed, the mean error is $0.45\% \pm 0.73\%$ on the finite-element data and $1.64\% \pm 0.30\%$ on the manufactured data, and the recovery then depends on the L-BFGS stage. A noise sweep with $N=5$ runs per level keeps the mean error at the percent level from 40~dB down to 0~dB, reaching $2.71\%$ at 0~dB, where the run-to-run spread is largest.

Within the limits of Sec.~\ref{sec:limitations}, the results show that exact constraints combined with a consistent non-dimensionalization identify a single piezoelectric coefficient from sparse, noisy field data more accurately and more reliably than penalty enforcement does, and to about one percent from 40~dB down to 10~dB. The full formulation also has no boundary or initial penalty weights to tune, which is where the adaptive baseline failed. Closing the gaps listed in Sec.~\ref{sec:limitations}, in particular boundary-driven excitation, measurable observations, and device-scale geometries, is the path toward in-situ characterization of piezoelectric transducers.

\section*{Data Availability}

The Python/PyTorch code and the data needed to reproduce the results of this work are available in the \texttt{Exact-Constrained-Piezo-PINN} repository on GitHub at \href{https://github.com/suhasbharadwajx/Exact-Constrained-Piezo-PINN}{https://github.com/suhasbharadwajx/Exact-Constrained-Piezo-PINN}. The repository has one folder for each study: \texttt{MMS-framework} for the manufactured-solution run, \texttt{FEM-framework} for the finite-element run, \texttt{SNR-sweep} for the noise sweep, and \texttt{Baseline-study} and \texttt{Ablation-study} for the soft-penalty baselines and the constrained variants of Secs.~\ref{sec:baseline} and~\ref{sec:ablation}, and \texttt{Extended-validation} for the displacement-only and field-accuracy runs of Sec.~\ref{sec:disp_only}.

\section*{Acknowledgements}

The author humbly thanks Prerana Ramkumar for the original idea that led to this work. During the preparation of this work, the author used Claude (Anthropic) to support code development and formatting (including the scripts released on GitHub), figure preparation, numerical cross-checks, and language editing. The author reviewed and verified all outputs, ran the experiments, and takes full responsibility for the content of the manuscript.

\section*{Conflict of Interest Statement}

The author declares no known competing financial interests or personal relationships that could have appeared to influence the work reported in this paper.

\bibliography{ref}

\end{document}